\documentclass[showpacs,prd,aps]{revtex4}

\pdfoutput=1

\usepackage{slashed}
\usepackage{hyperref}
\usepackage{amsmath}
\usepackage{amssymb}
\usepackage{latexsym}
\usepackage{amsfonts}
\usepackage{epsfig}
\usepackage{psfrag}
\usepackage{graphicx}

\def\hhref#1{\href{http://arxiv.org/abs/#1}{arXiv:#1}} % in bibliography

\usepackage{color}

\newcommand{\bea}{\begin{eqnarray}}
\newcommand{\ea}{\end{eqnarray}}
\newcommand{\eea}{\end{eqnarray}}

\begin{document}

\title{Complex Worldline Instantons and Quantum Interference in Vacuum Pair Production}

\author{Cesim~K.~Dumlu and Gerald~V.~Dunne}

\affiliation{Department of Physics, University of Connecticut,
Storrs CT 06269-3046, USA}

\begin{abstract}
We describe in detail a physical situation in which instantons are necessarily complex, not just Wick rotations of classical solutions to Euclidean spacetime.
These complex instantons arise in the semiclassical evaluation of vacuum pair production rates, based on Feynman's worldline path integral formulation. Even though the path integral is a sum over all real closed trajectories in spacetime, the semiclassical description of non-perturbative pair production is dominated by closed classical trajectories that are generically complex.
These closed trajectories contain segments associated with nonperturbative instanton suppression factors as well as segments producing phase factors that incorporate quantum interference effects. For a class of time-dependent electric fields we implement this procedure and demonstrate excellent quantitative agreement with alternative methods.

\end{abstract}

%\date{\today}


\maketitle

\section{Introduction}

The Heisenberg-Schwinger effect is the non-perturbative production of electron-positron pairs from the quantum electrodynamical (QED) vacuum under the influence of an external electric field \cite{sauter,he,schwinger,rafelski,fradkin}.
The general quantum field theoretic formalism for computing the associated probability was developed by Schwinger in terms of the QED effective action \cite{schwinger}. However, there are still serious obstacles to the implementation of a reliable quantitative computation when the external electric field is taken to be that for realistic high-intensity laser pulses.
Interest in this problem has been revived recently, spurred by new experimental developments in ultra-high intensity lasers \cite{tajima}.
Models of laser pulses with one-dimensional inhomogeneities, such as time-dependent linearly polarized electric fields are  well understood (although the question of pulse sequence optimization still stands), but the situation is much less clear for fields with multi-dimensional inhomogeneities, such as occur naturally in more realistic physical configurations of colliding high-intensity, spatially focussed, laser pulses \cite{dunne-eli}.
This is a pressing matter, because recent theoretical progress suggests \cite{Bulanov:2004de} that the critical peak field intensity required to observe this effect may in fact be several orders of magnitude lower than the estimate based on assuming a constant electric field \cite{sauter,he}, raising hopes that the effect may be observed experimentally in the not too distant future. In turn, this also raises important unresolved questions about back-reaction and cascading effects \cite{Bell:2008zzb}.

In the quantum field theoretic approach \cite{schwinger}, the technical  problem is to compute the non-perturbative imaginary part of the "effective action", $\Gamma[A]=-i\hbar \, \ln\, \det\left[ mc-i D\hskip -7pt /\, \right]
$, where the Dirac operator, $D\hskip -7pt / \equiv \gamma^\mu (\partial_\mu-i\frac{e}{\hbar c}A_\mu)$, defines the coupling between electrons and the applied (classical) electromagnetic field $A_\mu$ that represents the field produced by the laser pulse. The conventional approach to this problem in the case of a one-dimensional inhomogeneity  reduces it to a 1d scattering problem \cite{brezin,popov,nikishov}, invoking Feynman's picture of anti-particles as particles traveling backward in time \cite{feynman-positron}. There are then many possible approaches to compute pair production rates and the momentum spectra of the produced particles  \cite{brezin,popov,nikishov,gavrilov,kluger,schmidt,kimpage}. However, these one dimensional  methods do not generalize in a simple, efficient way to the multidimensional situation. There have been recent developments for multidimensional fields concerning finite-plane-wave fields \cite{Heinzl:2010vg}, and the numerical implementation \cite{Hebenstreit:2010vz} of the  Dirac-Heisenberg-Wigner formalism \cite{BialynickiBirula:1991tx}.
On the other hand, a natural semiclassical formulation of the general problem is in terms of worldline instantons, a semiclassical approximation to Feynman's worldline path integral expression for the QED effective action. This method has been quantitatively confirmed for certain one-dimensional field configurations, and the general formalism has been outlined for multi-dimensional field configurations \cite{Affleck:1981bma,wli,Dietrich:2007vw}. A technical obstacle to the implementation of the worldline instanton method in higher dimensions has been the physical interpretation of the complex classical trajectories that naturally arise. The purpose of this paper is to clarify the physical meaning of such complex classical trajectories, using a one-dimensional example for which we can confirm our results by comparison with other methods.

Usually instantons appear as solutions to the Euclidean classical equations of motion, in which $x^0\to x^4=i\, x^0$. In fact, this definition is too restrictive for the worldline picture,
and a more natural definition is to seek solutions with imaginary proper-time: $\tau\to s=i\tau$, as proposed by Rubakov {\it et al} \cite{rubakov}.
In simple text-book cases this transformation to imaginary proper-time goes hand-in-hand with the Wick rotation to imaginary  (Euclidean) time, but there are examples in which the spacetime instanton trajectories
$x^\mu(\tau)$ should be viewed as lying in complex Minkowski space \cite{rubakov,KeskiVakkuri:1996gn}.
In the case of QED, as studied here, the situation is even more interesting because the gauge coupling produces a Lorentz-force term in the relativistic classical equations of motion, $\ddot{x}_\mu=F_{\mu\nu}(x)\dot{x}^\nu$, which acquires a factor of "$i$" after rotating to imaginary proper time, so that the instanton equations are manifestly complex from the very beginning. [This is analogous to the effect of a magnetic field on a tunneling problem in non-relativistic quantum mechanics \cite{dykman}; it breaks time-reversal symmetry and makes the tunneling instanton equations complex.] In this paper we show that for the problem of QED vacuum pair production, complex instantons are needed to capture the physics of quantum interference between distinct instanton trajectories. This phenomenon of quantum interference arises for laser pulses with temporally localized electric field pulse shapes having sub-cycle structure, such as   ``carrier-envelope-phase'' or ``chirp'' features \cite{Hebenstreit:2009km,dd,Dumlu:2010vv}. In addition, appropriately chosen temporal sequences of  pulses can produce significant coherent enhancement in certain momentum modes, an explicit time-domain realization of multiple-slit interference \cite{Akkermans:2011yn}.
We treat both scalar and spinor QED to show explicitly how the interference terms are affected by the quantum statistics of the particles. We also note as motivation for studying complex instantons in QED the fact that complex trajectories are well-known in multi-dimensional tunneling phenomena in non-relativistic quantum mechanics \cite{miller,nakamura,aoyama,bezrukov,schiff}. Furthermore, the physical meaning of complex classical trajectories has recently been further elucidated by the study of PT-symmetry in quantum mechanics \cite{carl}.

In Section II we recall the worldline instanton formalism for the QED effective action, and explain why complex instanton solutions appear. In Section III we present the worldline instanton solution for the more general problem of finding the momentum spectrum for the produced electron-positron pair, and state the appropriate boundary conditions for finding the semiclassically important solutions of the complex classical equations of motion. Quantitative results are presented in Section IV, demonstrating excellent agreement with alternative methods of solution, and Section V contains our conclusions. An Appendix discusses an important and interesting numerical instability that occurs for certain ranges of values of the longitudinal momentum, and we present a simple resolution of this instability by taking advantage of the reparametrization invariance of the worldline path integral.

\section{Wordline Instanton Formalism}
\subsection{Worldline form of the QED effective action}

Following Schwinger \cite{schwinger}, we  compute the non-perturbative pair production probability $P$ from the  imaginary part of the effective action $\Gamma_{\text{eff}}[A]$ for the QED vacuum in a prescribed classical  background field $A_{\mu}(x)$:
\begin{eqnarray}
P=1-e^{-2\,\text{Im}\left[\Gamma_{\text{eff}}\right]/\hbar}
\approx \frac{2}{\hbar}\, \text{Im}\,\Gamma_{\text{eff}}\left[A\right]
\end{eqnarray}
For physically relevant configurations, $\text{Im}\,\Gamma_{\text{eff}}\left[A\right]/\hbar$ is extremely small, which justifies the approximation in the last step.
The effective action $\Gamma_{\text{eff}}\left[A\right]$ is defined, for spinor and scalar QED respectively, as (henceforth we set $\hbar=1$) \cite{schwinger}:
\begin{eqnarray}
\Gamma_{\rm eff}^{\rm spinor}[A]&=&-i \, \ln\, \det\left[ m c-i D\hskip -7pt / \hskip 3pt\right]=-\frac{i}{2} \,{\rm tr}\, \ln\left[m^2c^2+ D\hskip -7pt /\hskip2pt^2 \right]\\
\Gamma_{\rm eff}^{\rm scalar}[A]&=&i\, \ln\, \det\left[ m^2c^2+D_\mu^2\right]=i \,{\rm tr}\, \ln\left[ m^2c^2+D_\mu^2\right]
\label{effective}
\end{eqnarray}
The covariant derivative operator $D_\mu$ has been defined above, in the Introduction, and we adopt the space-time metric convention $g_{\mu\nu}={\rm diag}(1,-1,-1,-1)$.
Both Schwinger \cite{schwinger} and Feynman \cite{feynman} interpreted these effective actions in terms of quantum mechanical propagation in four-dimensional spacetime:
\begin{eqnarray}
\Gamma_{\rm eff}^{\rm spinor}[A]&=&\frac{i}{2}\int_0^\infty \frac{dT}{T} e^{-i \frac{m^2c^2}{2}\, T}\,{\rm tr}\,e^{-i\, {\mathcal H}_{\rm sp}\, T}\qquad, \quad {\mathcal H}_{\rm sp}=\frac{1}{2} D\hskip -7pt /\hskip2pt^2 \\
\Gamma_{\rm eff}^{\rm scalar}[A]&=&-i\int_0^\infty \frac{dT}{T} e^{-i \frac{m^2c^2}{2}\, T}\,{\rm tr}\,e^{-i\, {\mathcal H}_{\rm sc}\, T} \qquad, \quad {\mathcal H}_{\rm sc}=\frac{1}{2} D_\mu^2
\label{effective2}
\end{eqnarray}
The factor of $1/2$ in ${\mathcal H}$ is a convention \cite{feynman},  introduced by simple analogy with the form of the Hamiltonian in non-relativistic quantum mechanics. The integration variable $T$ can be thought of as the total propagation "time", which leads naturally  \cite{feynman} to a path integral expression for  the effective action. For scalar QED:
\begin{equation}
\Gamma_{\rm eff}^{\rm scalar}[A]=-i\int_0^{\infty}\frac{d T}{T} e^{-i \frac{m^2c^2}{2}\, T} \int d^4x\int_{x(0)=x(T)} \mathcal{D}x\,e^{-iS[x]}
\label{scalar}
\end{equation}
where $S$ is the classical action for a relativistic scalar charged particle, coupled to the gauge field $A_\mu(x)$, propagating around the closed trajectory $x^\mu(u)$ with a propagation period $T$:
\begin{eqnarray}
S[x^\mu(u); T]=
\int_0^T \left(\frac{1}{2}\frac{dx_\mu}{du}\frac{dx^\mu}{du} -\frac{e}{c} \frac{dx_\mu}{du}A^{\mu}(x)\right) du
\equiv \int_0^T L\left(x, \frac{dx}{du}\right)du
\label{action}
\end{eqnarray}
The  paths $x^\mu(u)$ are closed paths in four-dimensional spacetime, parametrized by a  parameter $u$, which we relate to proper-time in the following subsection. For spinor QED there is an additional spin interaction, because $D\hskip -7pt /\hskip2pt^2=D_\mu^2-\frac{e}{2c}\sigma^{\mu\nu}F_{\mu\nu}$, and the effective action for spinor QED can be written with an additional Grassmann path integration \cite{chris,polyakov,kleinert}:
\begin{equation}
\Gamma_{\rm eff}^{\rm spinor}[A]=\frac{i}{2}\int_0^{\infty}\frac{d T}{T} e^{-i \frac{m^2c^2}{2}\, T} \int d^4x\int_{x(0)=x(T)} \mathcal{D}x\,e^{-iS[x]}\int\mathcal D \psi\, e^{-i\int_0^T\left(i\psi^\mu\frac{d\psi_\mu}{du}-i \frac{e}{c} \psi_\mu F^{\mu\nu}(x)\psi_\nu\right)du}
\label{spinor}
\end{equation}
where $S$ is as in (\ref{action}).

\subsection{Semiclassical approximation: worldline instantons}

The path integrals in (\ref{scalar}) and (\ref{spinor}) are of course only analytically calculable in special idealized cases, so we must resort to approximation methods.
For non-pertubative questions such as the the pair production probability one can use the  numerical worldline approach \cite{Gies:2001tj}, or a semiclassical evaluation based on worldline instantons \cite{Affleck:1981bma,wli}. In this paper we follow the worldline instanton approach, in which we search first for a saddle point solution to the bosonic path integral by solving the classical equations of motion for relativistic motion of  a charged spinless particle:
\begin{equation}
\frac{d^2x^{\mu}}{du^2}=\frac{e}{c}\,F^{\mu\nu}(x)\frac{dx_{\nu}}{du}
\label{eom}
\end{equation}
A closed trajectory solution to these classical equations is called a "worldline instanton". In certain physical situations, such as for non-perturbative pair production from vacuum, these classical solutions give a dominant contribution to the path integral in (\ref{scalar}) and (\ref{spinor}).

The classical equations (\ref{eom}) have an obvious first integral,  the "energy" $H=p_\mu\frac{dx^\mu}{du}-L=\frac{1}{2}\frac{dx_\mu}{du}\frac{dx^\mu}{du}$, which is a constant of motion. This constant is fixed by making also a saddle-point approximation to the $T$ integral, which gives a critical condition:
\begin{eqnarray}
\frac{m^2 c^2}{2}+\frac{\partial S}{\partial T}=0
\label{legendre}
\end{eqnarray}
Since the variation of the action with respect to the period $T$, namely $\frac{\partial S}{\partial T}$, is equal to minus the conserved energy, this implies the normalization
\begin{eqnarray}
\frac{dx_\mu}{du}\frac{dx^\mu}{du}=m^2 c^2
\label{proper}
\end{eqnarray}
Using the relation, $\frac{dt}{d\tau}=1/\sqrt{1-\vec{v}^2/c^2}$, between time $t=x^0/c$ and proper-time $\tau$,  this identifies the propagation parameter $u$ as a multiple of the proper-time:
\begin{eqnarray}
u=\frac{\tau}{m}
\label{pt}
\end{eqnarray}
We can therefore identify $m T$ with the total proper-time of the quantum mechanical evolution in (\ref{effective2}).
[We shall see later, in the Appendix, that a different scaling of the proper-time evolution parameter leads to some numerical advantages in certain situations.]

The critical saddle point period $T_c$ is determined by the condition (\ref{legendre}), and when we evaluate the full exponent of the $T$ integral in (\ref{scalar}) we obtain Hamilton's characteristic function $W[x^\mu(u); \frac{1}{2}m^2c^2]=S[x^\mu(u); T]+\frac{1}{2}m^2c^2 T$. We will refer to this as
the classical action for the motion of the relativistic charged particle to traverse its closed trajectory with a prescribed "energy", equal to $\frac{1}{2}m^2c^2$. This is just the familiar Legendre transform of classical mechanics, relating the action $S[x; T]$ and the characteristic function $W[x; {\mathcal E}]=S[x; T]+{\mathcal E}T$, which implies: $\frac{\partial S}{\partial T}=-{\mathcal E}$, and $\frac{\partial W}{\partial {\mathcal E}}=T$. Here, in this relativistic problem the role of the "energy" ${\mathcal E}$ is played by $\frac{1}{2}m^2c^2$, in complete agreement with Feynman's interpretation of the Klein-Gordon and Dirac equation in terms of proper-time \cite{feynman} (based on results of Fock and Nambu \cite{fock,nambu}). Thus, on the classical solution, the characteristic function becomes
\begin{eqnarray}
W[x^\mu(u); \frac{1}{2}m^2c^2]=
\int_0^T \left(\frac{1}{2}\frac{dx_\mu}{du}\frac{dx^\mu}{du} -\frac{e}{c} \frac{dx_\mu}{du}A^{\mu}(x)+\frac{1}{2}m^2 c^2\right) du
\equiv \int_0^T p_\mu \frac{dx^\mu}{du}du
\label{w}
\end{eqnarray}
This is the classical function that appears in the exponent after making a semiclassical approximation. Note that there may be several classical saddle point trajectories $x^\mu(u)$ relevant to the semiclassical approximation.

\subsection{The need for complex worldline instantons}

The worldline instanton approach was originally suggested for  QED with a constant background electric field \cite{Affleck:1981bma}, then extended to QED in background fields with one-dimensional inhomogeneities in \cite{wli}, and a general formalism was also proposed for more general fields, based on an analogy with Gutzwiller's trace formula \cite{Dietrich:2007vw}.
The most difficult part of the computation is to find the semiclassically important classical trajectories. The first observation is that we seek {\it closed} trajectory solutions; this is because the effective action involves a trace, so that the quantum mechanical path integral in (\ref{scalar}, \ref{spinor}) is expressed as a sum over {\it closed paths} in four-dimensional spacetime. But we still need to specify some initial conditions in order to search for appropriate saddle point solutions. Previously it had been suggested to look for closed trajectory solutions to the {\it Euclidean} classical equations of motion \cite{wli,Dietrich:2007vw}, obtained from (\ref{eom}) by a Wick rotation, $x^0\to x^4=i\,x^0$. We show here that this prescription needs to be refined and extended, in order to describe quantum interference effects.

We begin with the formalism of Rubakov et al \cite{rubakov,KeskiVakkuri:1996gn,bezrukov},  that an instanton solution is associated with a deformation of the  contour of the $T$ integral onto the imaginary axis, so that we look for solutions with imaginary "proper-time" parameter: $u\to s=i\,u$. In the familiar case of scalar [non-derivative] couplings, the  classical equations of motion acquire a sign change:
\begin{eqnarray}
\frac{d^2 x^\mu}{du^2}=-\frac{\partial V(x)}{\partial x_\mu} \quad \rightarrow\quad
\frac{d^2 x^\mu}{d s^2}=+\frac{\partial V(x)}{\partial x_\mu}
\label{scalar-imag}
\end{eqnarray}
Note that the equations of motion  remain real under this operation. On the other hand, for the gauge coupling of the QED case, imaginary proper-time introduces a factor of "$i$" into the classical equations of motion.
\begin{eqnarray}
\frac{d^2 x^\mu}{du^2}=\frac{e}{c}\,F^{\mu\nu}(x)\frac{d x_\nu}{du} \quad \rightarrow\quad
\frac{d^2 x^\mu}{d s^2}=i\,\frac{e}{c}\,F^{\mu\nu}(x)\frac{d x_\nu}{d s}
\label{qed-imag}
\end{eqnarray}
Therefore, the instanton equations are generically complex, and so the solutions will be generically complex, as will be the classical action evaluated on such a solution. This raises the  question: what is the physical significance of such complex classical solutions? We answer this question in the remainder of this paper.

\begin{figure}[htb]
\includegraphics[scale=0.65]{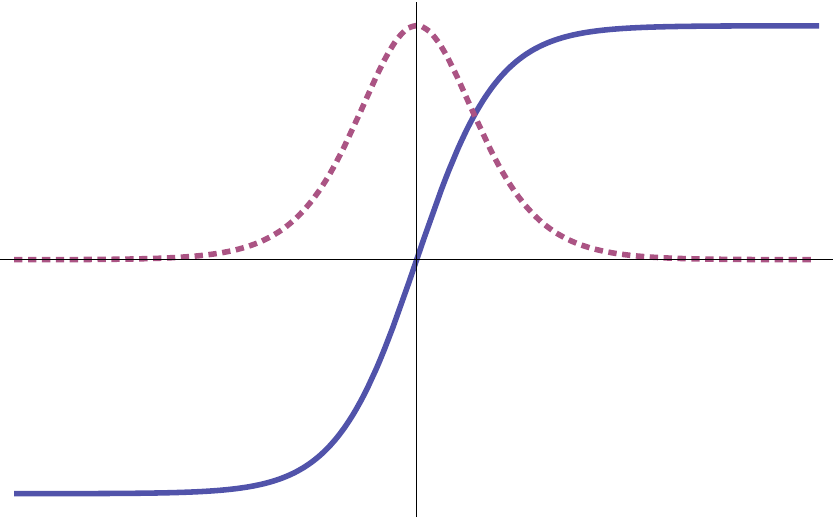}\quad
\includegraphics[scale=0.65]{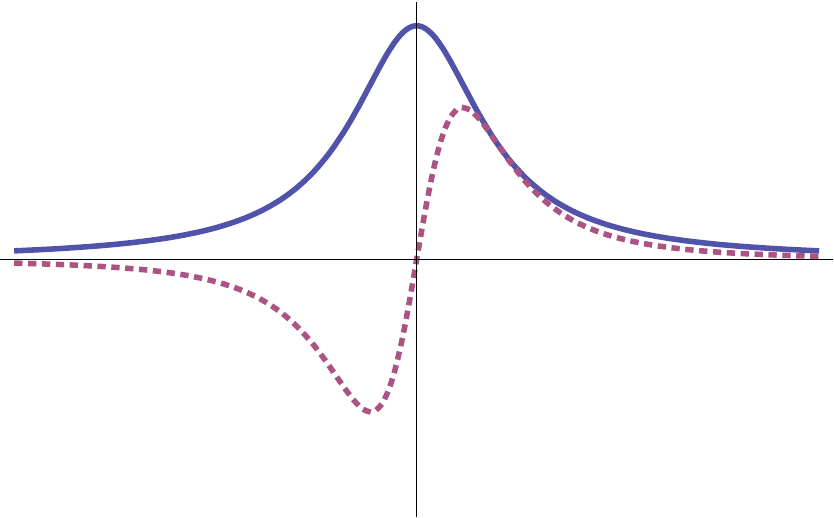}\quad
\includegraphics[scale=0.65]{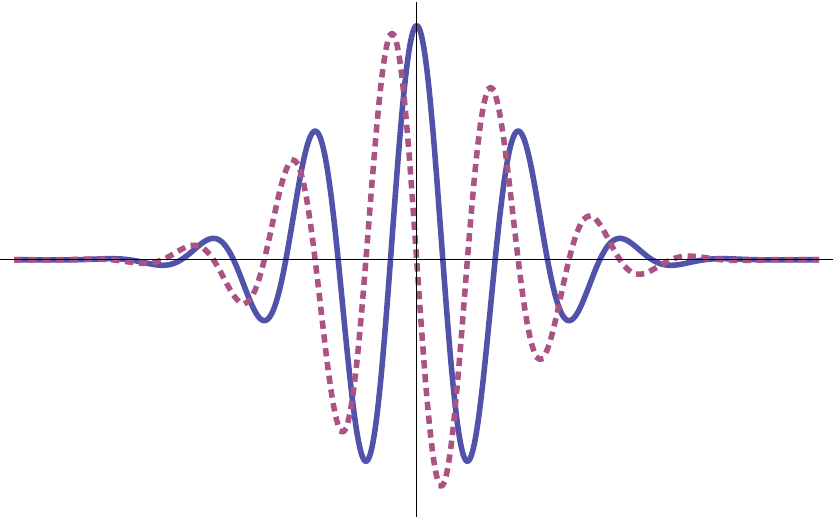}
\caption{Sketch of basic shapes of gauge fields [solid, blue curves] $A_3(x^0)$ and the corresponding electric field [dotted, red, curves] $E_3(x^0)$. In the first case, the gauge field is a monotonic odd function of time and the electric field is a single localized pulse. In this case there are no quantum interference effects. For the next two cases, the gauge field is an even function of time, the electric field is an odd function of time,  and there are significant quantum interference effects, as discussed in \cite{dd} using WKB.}
\label{fig1}
\end{figure}

An important comment concerning previous work on worldline instantons \cite{Affleck:1981bma,wli}, as well as on early work on the WKB scattering approach \cite{brezin,popov,nikishov}, is that they concentrated on two particular sub-classes of problems. The first class consists of time-dependent electric fields linearly polarized along a particular spatial axis [say, the $x^3$ axis], so that the gauge field can be written with only one non-zero component $A_3(x^0)$. Classic cases are: $A_3(x^0)=E x^0$, for a uniform field; $A_3(x^0)=E/\omega \sin(\omega x^0)$ for a monochromatic sinusoidal field; $A_3(x^0)=E/\omega \tanh(\omega x^0)$ for a temporally localized single-pulse field. All these examples have the important property that $A_3(x^0)$ is an {\it odd function of time}. The second class of fields involves static but spatially inhomogeneous fields represented by a scalar potential $A_0(\vec{x})$. Now, observe that in each of these cases, the complex classical equations of motion in (\ref{qed-imag}) reduce again to real equations if we combine the analytic continuation, $u\to s=i\,u$, with the Wick rotation, $x^0\to x^4=i\,x^0$, and $A^0\to A^4=i\, A^0$. In the former case, this is only true if $A_3(x^0)$ is an odd function. Otherwise, the equations remain complex. This explains why the previous analyses were able to produce consistent and correct results using as worldline instanton equations the classical {\it Euclidean} equations of motion with imaginary proper-time.

As a simple illustrative example, consider the case of a constant electric field, of strength $E$. Then the classical trajectories are  hyperbolic paths: $x^0(u)=\frac{mc^2}{e E} \sinh(e E u/c)$, and $x^3(u)=\frac{mc^2}{e E} \cosh(e E u/c)$. These are not periodic closed paths, but we can make them closed and periodic if we take $u\to s=i\,u$. We then obtain complex solutions,  $x^0=-i\,\frac{mc^2}{e E}  \sin(e E s/c)$, and $x^3=\frac{mc^2}{e E} \cos(e E s/c)$, which become real again when expressed in terms of the Euclidean time $x^4=i\, x^0$. But the reality of the worldline instanton solution in Euclidean spacetime  is an "accident", a direct result of $A_3(x^0)$ being an odd function of $x^0$. For this solution, the period is $T_c=\frac{2\pi c}{eE}$,  and we can evaluate the characteristic function (\ref{w}) as:
\begin{eqnarray}
W[x^\mu(u); \frac{1}{2}m^2 c^2]&=&\int_0^{\frac{2\pi c}{eE}}\left(p_0\frac{dx^0}{ds}+p_3\frac{dx^3}{ds}\right)ds
\nonumber\\
&=&i\int_0^{\frac{2\pi c}{eE}}\left(\frac{dx^0}{ds}\right)^2 ds
\nonumber\\
%&=&-i m^2 c^2\int_0^{\frac{2\pi c}{eE}}\cos^2\left(\frac{e Es}{c}\right) ds
%\nonumber\\
&=&-i\frac{m^2c^3\pi}{eE}
\label{constant}
\end{eqnarray}
Then  the semiclassical approximation to the scalar QED effective action leads to
\begin{eqnarray}
\text{Im}\, \Gamma_{\rm eff}^{\rm scalar}\approx {\mathcal P}\, e^{-i W_{\rm instanton}}\approx {\mathcal P}\, e^{-\frac{m^2 c^3\pi}{eE}}
\end{eqnarray}
which is the familiar result of Sauter \cite{sauter}, Euler and Heisenberg \cite{he} and Schwinger \cite{schwinger}, and
${\mathcal P}$ is a simple prefactor. Other examples in which the explicit worldline instanton trajectories and actions  can be evaluated in closed form are given in \cite{wli}, with results in agreement with the corresponding WKB treatment \cite{brezin,popov} (including also the prefactors).

In more realistic time-dependent electric fields, such as those having an envelope structure as well as an oscillatory structure, there are quantum interference effects, which can produce both enhancement and suppression \cite{Hebenstreit:2009km}. These cases are associated with vector potentials that cannot be written as an odd function of time, as sketched in Figure \ref{fig1}. The semiclassical analysis of such systems, incorporating quantum interference, has been given in \cite{dd} using the WKB approach. Here we explain how to solve these quantum interference problems using the more general formalism of worldline instantons. Since the worldline instanton equations remain manifestly complex,  we confront directly the problem of the physical meaning of complex instantons. Our motivation for studying this class of problems first is that we have results with which to compare, so that we can quantitatively verify the validity of our approach.

There may be several classical saddle point trajectories, $x^\mu_{(j)}(u)$ , labeled by an index $(j)$, and the imaginary part of the effective action is then approximated by
\begin{eqnarray}
{\rm Im}\,\Gamma_{\rm eff}^{\rm scalar}\approx\sum_j\mathcal P^{(j)}\,e^{-i \,W_{\rm instanton}^{(j)}}
\label{approx1}
\end{eqnarray}
where $W_{\rm instanton}^{(j)}$ is the characteristic function (\ref{w}), evaluated on the $j^{\rm th}$ saddle point solution, and $\mathcal P^{(j)}$ is a (readily calculable) prefactor. In the case of spinor QED, the saddle point trajectory is the same, but there is an additional spin factor coming from the evaluation of the (Gaussian) spinor path integral in (\ref{spinor}), evaluated on the critical trajectory $x_{\rm cl}^{(j)}(u)$ \cite{kleinert}:
\begin{eqnarray}
{\rm Im}\,\Gamma_{\rm eff}^{\rm spinor}\approx - \sum_j\mathcal P^{(j)}\, {\rm det}^{1/2}\left(\delta_{\mu\nu}\frac{d}{du}-i\frac{e}{mc}F_{\mu\nu}(x_{\rm cl}^{(j)}(u))\right)\,e^{- i\,W_{\rm instanton}^{(j)}}
\label{approx2}
\end{eqnarray}
The determinant spin factor can be computed straightforwardly using the Gelfand-Yaglom method, since it only involves ordinary differential operators, as can the prefactors  \cite{wli,Dietrich:2007vw,kleinert}.

\section{Momentum spectra for particles produced in time-dependent electric fields}

\subsection{General formalism}

In this section we extend the worldline instanton method of \cite{Affleck:1981bma,wli} to compute not only the total probability of pair production, but also the momentum spectrum of the produced particles. In the situation where the vector potential $A_3(x^0)$ is a function only of $x^0$, the spatial momenta of the electron-positron pair, $\vec{p}=(p_\perp, p_3)$, are good quantum numbers and can be used to characterize the final states. Furthermore,  in a strong field the pair production is predominantly along the direction of the electric field, so we can neglect $p_\perp$ and concentrate on the dependence of the number of produced pairs on the longitudinal momentum $p_3$. Studies of temporally structured electric field pulses have revealed an intricate dependence on $p_3$, due to quantum interference effects \cite{Hebenstreit:2009km,dd}. In this Section we show that the analysis of this momentum dependence using worldline instantons requires {\it complex} worldline instanton trajectories, not simply Euclidean classical trajectories.

To address the momentum dependence we convert the worldline path integral expressions (\ref{scalar}, \ref{spinor}) for the QED effective action into phase space path integral expressions. Since the only difference between spinor and scalar QED in the semiclassical approximation is the spin factor determinant in (\ref{approx2}), we first concentrate on the scalar QED case in order to find the instanton trajectories.
\begin{equation}
\Gamma_{\rm eff}^{\rm scalar}[A]=-i\int_0^{\infty}\frac{d T}{T}e^{-i\frac{m^2c^2}{2}\, T}\int d^4x\int_{x(0)=x(T)} \mathcal{D}x\, \int\mathcal{D}p\, e^{-i\int_0^T\left(p_\mu\frac{dx^\mu}{du}-H(x, p)\right)}
\label{scalar-phase}
\end{equation}
where the classical Hamiltonian density is
\begin{eqnarray}
H(x, p)&=&p_\mu \frac{dx^\mu}{du} -L(x, \dot{x})=\frac{1}{2}\left(p_\mu+\frac{e}{c} A_\mu(x)\right)^2
\label{ham}
\end{eqnarray}
Since the external field $A_\mu$ is independent of the spatial coordinates $\vec{x}$, the spatial path integral can be done, producing delta functions in $\frac{d\vec{p}}{du}$, thus imposing the conservation of spatial momentum. This means that the functional integrals for the spatial momentum reduce to ordinary integrals: $\int \mathcal D^3p\to \int d^3p$. Then we can convert the remaining phase space path integral over $x^0$ and $p^0$ back to a coordinate space path integral, leading to a worldline path integral expression  in terms of a single coordinate $x^0(u)$,
parametrized by the spatial momenta $\vec{p}$:
\begin{equation}
\Gamma_{\rm eff}^{\rm scalar}[A]=-i\left(V_3 \int d^2p_\perp\int dp_3\right) \int_0^{\infty}\frac{d T}{T}e^{-i\frac{m^2c^2}{2}\, T} \int dx^0 \int_{x^0(0)=x^0(T)} \mathcal{D}x^0\,  e^{-iS[x^0]}
\label{x0}
\end{equation}
where the classical action is:
\begin{eqnarray}
S[x^0(u)]=\int^T_0 \left(\frac{1}{2}\left(\frac{dx^0}{du}\right)^2+\frac{1}{2}\left(p_3+\frac{e}{c}A_3(x^0)\right)^2+\frac{p_\perp^2}{2}\right)du
\label{x0action}
\end{eqnarray}
Recalling the scaling (\ref{pt}) between between $u$ and proper-time $\tau$, this effective action expression (\ref{x0}, \ref{x0action}) has the form of a quantum mechanical path integral in the single (time) coordinate $x^0(\tau)$, parametrized by the proper-time $\tau$, with a "potential", $V(x^0)=-\frac{1}{2}\left(p_3+\frac{e}{c}A_3(x^0)\right)^2-\frac{p_\perp^2}{2}$, that depends parametrically on the spatial momenta $p_3$ and $p_\perp$. This agrees completely with the WKB picture of pair production for time dependent electric fields as a quantum mechanical Schr\"odinger over-the-barrier scattering problem in the time-domain \cite{brezin,popov,dd}.
The argument of the $\vec{p}$ integral in (\ref{x0}) has a non-perturbative imaginary part that gives the probability of producing pairs with spatial momentum $\vec{p}$.

\subsection{Semiclassical approximation for momentum spectrum}

We now make a semiclassical approximation to the effective action in (\ref{x0}, \ref{x0action}). First, we neglect the transverse momenta, setting $p_\perp=0$. This only affects prefactor terms, and can easily be incorporated if desired, and will not be important in what follows. Then the classical equation of motion
is
\begin{eqnarray}
\frac{d^2x^0}{du^2}&=& \frac{e}{c}\left(p_3+\frac{e}{c}A_3(x^0)\right) \partial_0 A_3(x^0)
\label{x0-eom}
\end{eqnarray}
This equation (\ref{x0-eom}) is of course just the remaining nontrivial classical equation of motion from (\ref{eom}) after integrating the $x_3$ equation to give $\frac{dx_3}{du}=p_3+\frac{e}{c}A_3(x^0)$, with $p_3$ arising as an integration constant. In \cite{wli}, in computing the total pair production rate rather than the momentum dependence, this integration constant was taken to vanish, with a Gaussian momentum integration producing certain prefactor terms that were explicitly computed and shown to agree with the WKB result \cite{popov}. Here we retain the $p_3$ dependence in the equation of motion (\ref{x0-eom}) in order to find the longitudinal momentum spectrum of the produced particles. Thus, the solution $x^0(u)$ will depend parametrically on $p_3$, as will the classical action when evaluated on that classical solution:
\begin{eqnarray}
S[x^0(u); T]= \int_0^T \left(\left(\frac{dx^0}{du}\right)^2-\frac{1}{2}m^2 c^2\right) du
%-\frac{1}{2} mc^2\right)
\label{classical}
\end{eqnarray}
Here we have set $p_\perp=0$ and used the proper-time  relation
\begin{eqnarray}
c^2(\dot{x}^0)^2=c^2+\frac{1}{m^2}\left(p_3+\frac{e}{c}A_3(x^0)\right) ^2
\label{v}
\end{eqnarray}
which also expresses the existence of a first integral for the equation of motion (\ref{x0-eom}). The term $-\frac{1}{2}m^2 c^2 T$ in (\ref{classical}) cancels against a similar term in the $T$ integral in (\ref{x0}), and so the resulting exponent is the Hamilton characteristic function, the action for a closed trajectory of fixed "energy" $\frac{1}{2} m^2 c^2$:
\begin{eqnarray}
W[x^0(u); \frac{1}{2}m^2c^2]= \int_0^T \left(\frac{dx^0}{du}\right)^2 du
%-\frac{1}{2} mc^2\right)
\label{reduced}
\end{eqnarray}
Then the imaginary part of the QED effective action, in scalar and spinor QED respectively, is given by:
\begin{eqnarray}
{\rm Im}\,\Gamma_{\rm eff}^{\rm scalar}\approx \int dp_3\sum_j\mathcal P^{(j)}\,e^{- i\, W_{\rm instanton}^{(j)}(p_3)}
\label{approx3}
\end{eqnarray}
\begin{eqnarray}
{\rm Im}\,\Gamma_{\rm eff}^{\rm spinor}\approx - \int dp_3\sum_j\mathcal P^{(j)}
\,{\rm det}^{1/2}\left(\delta_{\mu\nu}\frac{d}{du}-i\frac{e}{mc}F_{\mu\nu}(x_{\rm cl}^{(j)}(u; p_3)) \right)
\, e^{-i\, W_{\rm instanton}^{(j)}(p_3)}
\label{approx4}
\end{eqnarray}
where the sum is over all relevant semiclassical trajectories (to be specified explicitly in the next subsection).

\subsection{Boundary conditions for the worldline instanton trajectories}

\begin{figure}[htb]
\includegraphics[scale=.8]{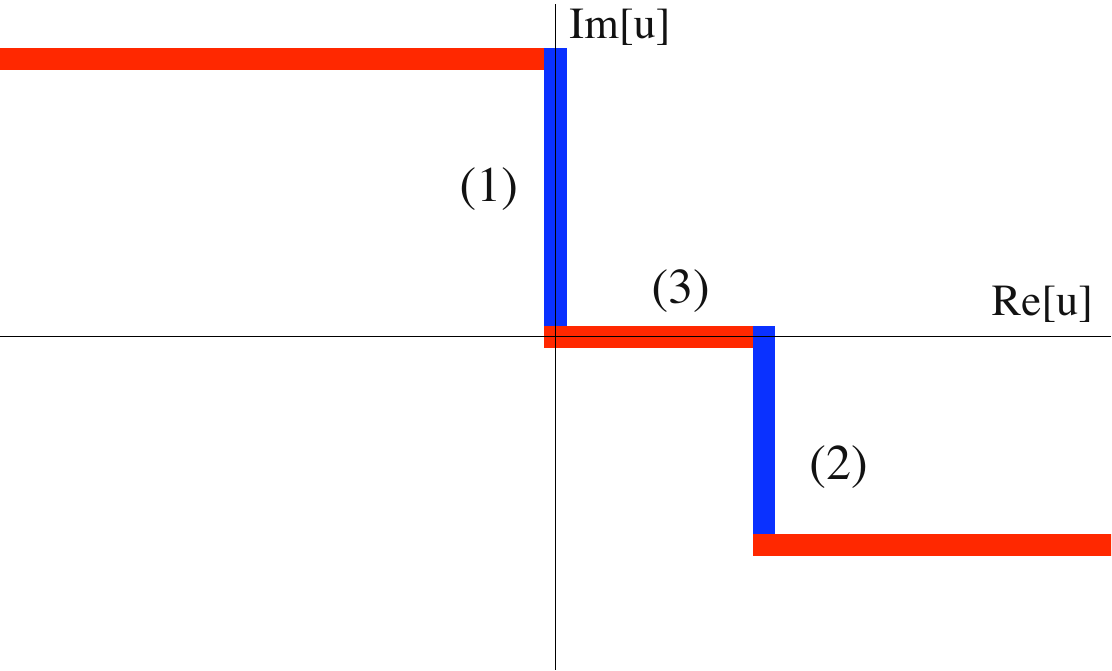}
\caption{Sketch of the complex contour in the complex $u$ plane, showing two distinct "instanton segments", labeled (1) and (2), having constant real part of $u$, and an "interference segment", labeled (3), having constant imaginary part of $u$, and connecting the two distinct instanton segments.}
\label{fig2}
\end{figure}

We now specify the appropriate boundary conditions for finding solutions to the classical equations of motion (\ref{x0-eom}). To find instanton solutions we expect to take $u$ (and hence $T$) to be imaginary. This corresponds to a deformation of the contour of integration of the $T$ integral \cite{rubakov}. This then leads to (some) closed path trajectories. However, for background gauge fields with nontrivial temporal structure, there can be more than one different instanton solution, and we need to be able to find all relevant instanton trajectories and tie them together. Thus, we consider a contour for $u$ that can have "vertical" segments along which the real part is constant, which we refer to as instanton segments, and in addition we  consider "horizontal" segments along which the imaginary part of $u$ is constant, which we refer to as "interference" segments. This situation is illustrated in Figure \ref{fig2}. This leads to a difference  of sign in the equation of motion (\ref{x0-eom}) for the two types of solution.

Next we specify the initial conditions.
As mentioned at the end of Section IIIA, the momentum spectrum problem is equivalent to the WKB scattering problem treated in \cite{dd}.
Motivated by this WKB analysis, we propose the boundary condition that the classical worldline trajectories should begin and end at WKB turning points, which are defined as points where
\begin{eqnarray}
m^2 c^2+\left(p_3+\frac{e}{c}A_3(x^0)\right) ^2=0
\label{tps}
\end{eqnarray}
These points lie in the complex $x^0$  plane, and  they occur in complex conjugate pairs in the physically relevant case where $A_3$ is a real function of its argument. The turning points move around in the complex plane as $p_3$ varies, but always remain in complex conjugate pairs.
Given this  initial condition for $x^0$, then because of (\ref{v}), the corresponding initial condition is that  $\frac{dx^0}{du}$ must vanish.
Thus we are led to the initial condition for our semiclassical trajectories: that the initial "velocity" vanishes, $\frac{dx^0}{du}=0$, and the initial (and final) point is a turning point. We therefore seek solutions as follows:
\begin{itemize}

\item {\bf Instanton segments:}
For an instanton segment, we take a complex proper-time evolution parameter $u$ with constant real part, $\text{Re}(u)=\text{constant}$, corresponding to one of the "vertical" segments in the contour depicted in Figure \ref{fig2}. Therefore, we have the following equations of motion to solve:
\begin{eqnarray}
\frac{d^2x^0}{du^2}&=&-\frac{e}{c} \left(p_3+\frac{e}{c}A_3(x^0)\right) \partial_0 A_3(x^0)
\label{x0-instanton-a}\\
x^0(u_{\rm initial})&=&x^0_{(j)} \qquad , \quad \text{a turning point solution of (\ref{tps})}
\label{x0-instanton-b}\\
\left. \frac{dx^0}{du}\right |_{u=u_{\rm initial}}&=&0
\label{x0-instanton-c}
\end{eqnarray}

\item {\bf Interference segments:}
For an interference segment, we take a complex proper-time parameter $u$ with constant imaginary part, $\text{Im}(u)=\text{constant}$, corresponding to one of the "horizontal" segments in the contour depicted in Figure \ref{fig2}. Therefore, we have the following equations of motion to solve:
\begin{eqnarray}
\frac{d^2x^0}{du^2}&=&+\frac{e}{c} \left(p_3+\frac{e}{c}A_3(x^0)\right) \partial_0 A_3(x^0)
\label{x0-interference-a}\\
x^0(u_{\rm initial})&=&x^0_{(j)} \qquad , \quad \text{a turning point solution of (\ref{tps})}
\label{x0-interference-b}\\
\left. \frac{dx^0}{du}\right |_{u=u_{\rm initial}}&=&0
\label{x0-interference-c}
\end{eqnarray}

\end{itemize}
Notice the different sign in the equations of motion for the two types of segment.

\begin{figure}[htb]
\includegraphics[scale=0.7]{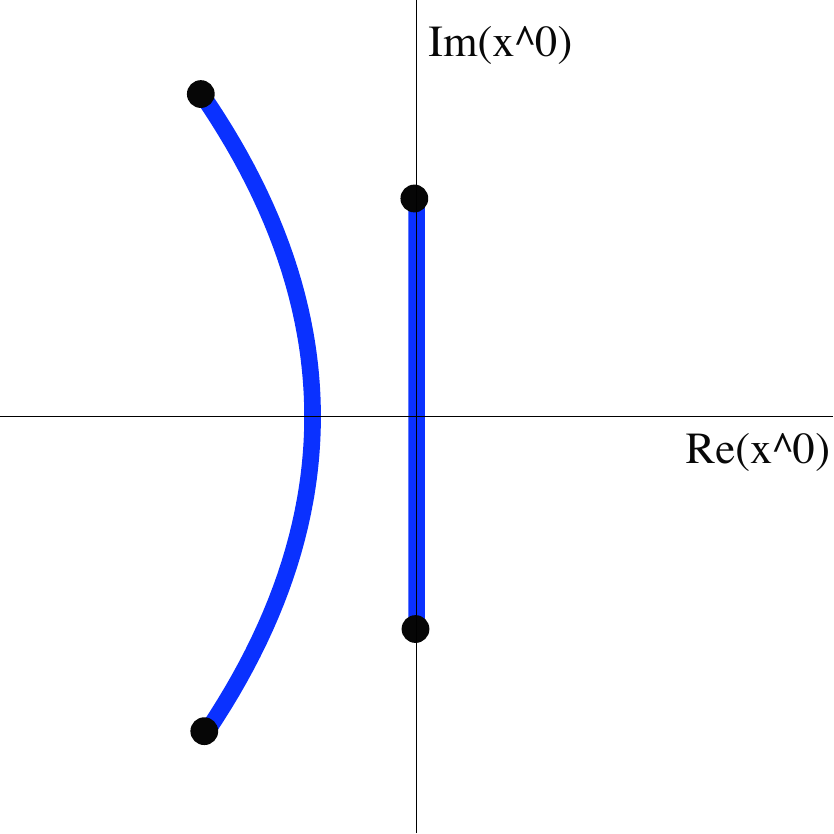}\hskip 3cm
\includegraphics[scale=0.7]{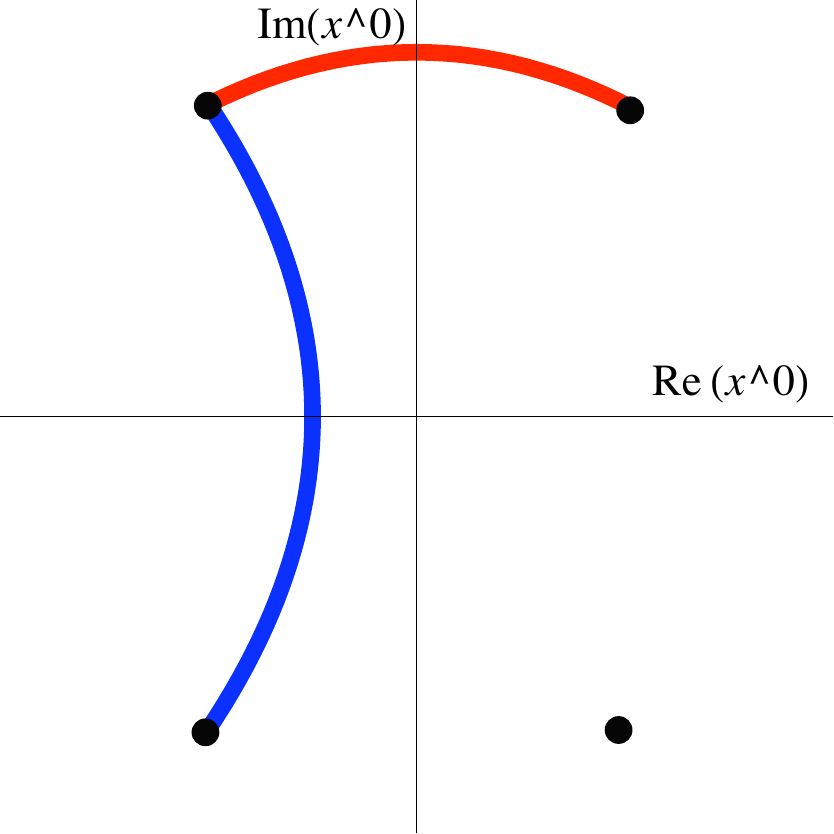}
\caption{Sketch of basic shapes of worldline instanton trajectory segments. The blue segments are instanton segments and the red ones are interference segments. In the first plot, we show the turning point pair for two different values of $p_3$. When $p_3=0$ the pair lie on the imaginary $x^0$ axis, and the instanton trajectory lies entirely on the imaginary $x^0$ axis. This is the usual Wick rotation to imaginary time. The other curve shows the situation for $p_3\neq 0$, in which case the turning points lie off the imaginary axis [note they still form a complex conjugate pair], and the instanton trajectory connecting them is curved. So this solution is not captured by a Wick rotation. In the second plot, we show the situation for a field with four turning points, in two complex conjugate pairs. Here, even for  $p_3=0$, the turning points lie off the imaginary axis, and there are two types of trajectories. The blue instanton trajectory connects a complex conjugate pair of turning points, while the red interference trajectory connects two different turning points, not a complex conjugate pair.
}
\label{fig3}
\end{figure}

The important observation is that these trajectories go from one turning point to another. For an instanton segment the trajectory goes from one turning point to its complex conjugate, while for an interference segment the trajectory goes from one turning point to another distinct turning point with different real part [if the field $A_3(x^0)$ is such that there is another such distinct turning point]. This is illustrated in Figure \ref{fig3}. In Figure \ref{fig3}a  we see two different types of instanton segments for the case $A_3(x^0)=E/\omega \tanh(\omega x^0)$. When $p_3=0$, the turning points lie on the imaginary $x^0$ axis, and the instanton trajectory goes along the imaginary axis, connecting the turning point to its complex conjugate. This explains why the Wick rotation to imaginary time, $x^4=i\, x^0$, is sufficient for this case \cite{wli}. However, when $p_3\neq 0$, two things change: first, the turning points move off the imaginary axis into the complex plane, and second, the trajectory is no longer linear. Thus, in this case with $p_3\neq 0$ we must consider truly complex instanton trajectories for $x^0(s)$, even though for this vector potential there are no interference trajectories (reflecting the simple single-bump structure of the corresponding electric field).

The second example, in Figure \ref{fig3}b, is a case in which interference effects do arise, taking the example $A_3(x^0)=E/\omega/(1+ (\omega x^0)^2)$, that was studied in \cite{dd} using WKB methods. Here, there are four turning points, in two complex conjugate pairs, and they never all lie on the imaginary $x^0$ axis. Thus, the naive Wick rotation to imaginary time is not sufficient for this problem. Nevertheless, using the prescription described above we find both instanton and interference trajectories, as illustrated in Figure \ref{fig3}b. There are instanton segments connecting complex conjugate turning points, and interference trajectories connecting the two distinct turning points having different real parts. As $p_3$ changes, all four turning points move, and the shape of the trajectories change, but the pattern remains the same.

\begin{figure}[htb]
\includegraphics[scale=0.5]{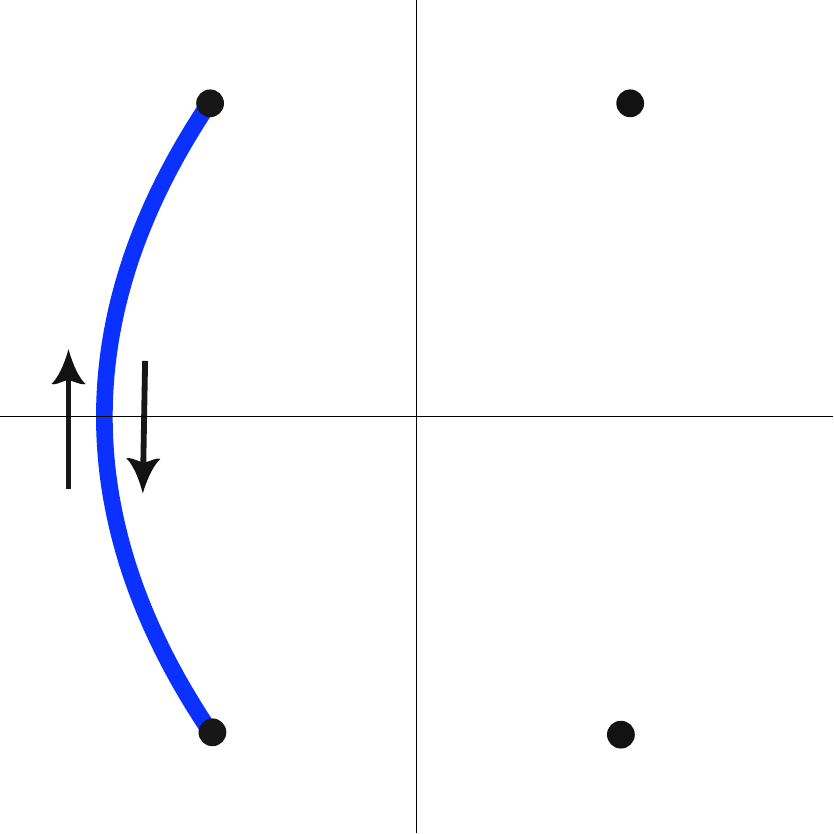}\quad
\includegraphics[scale=0.5]{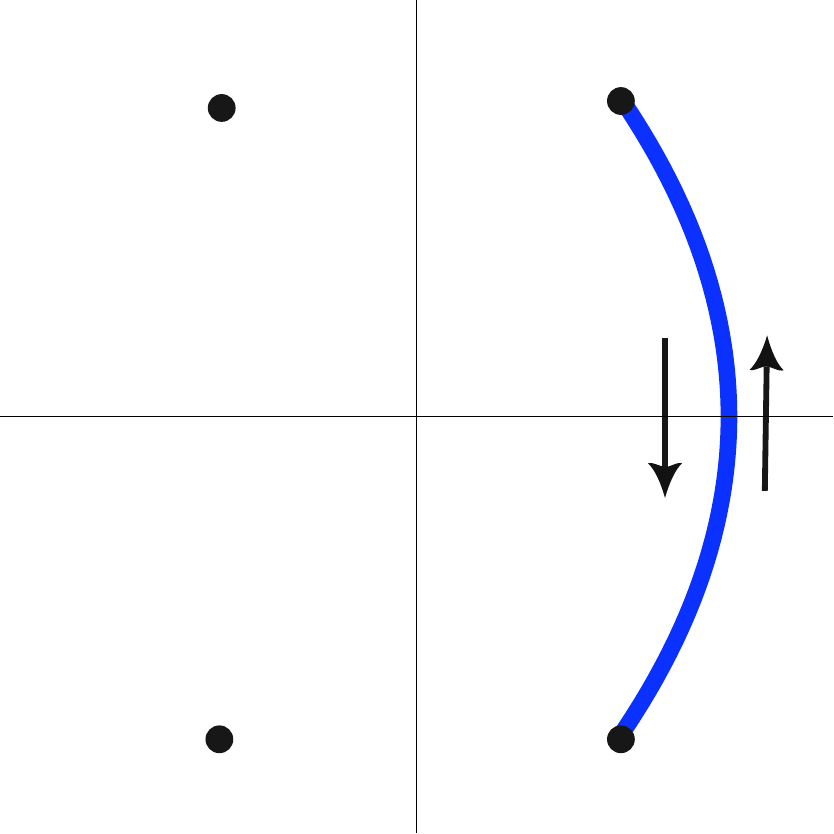}\quad
\includegraphics[scale=0.5]{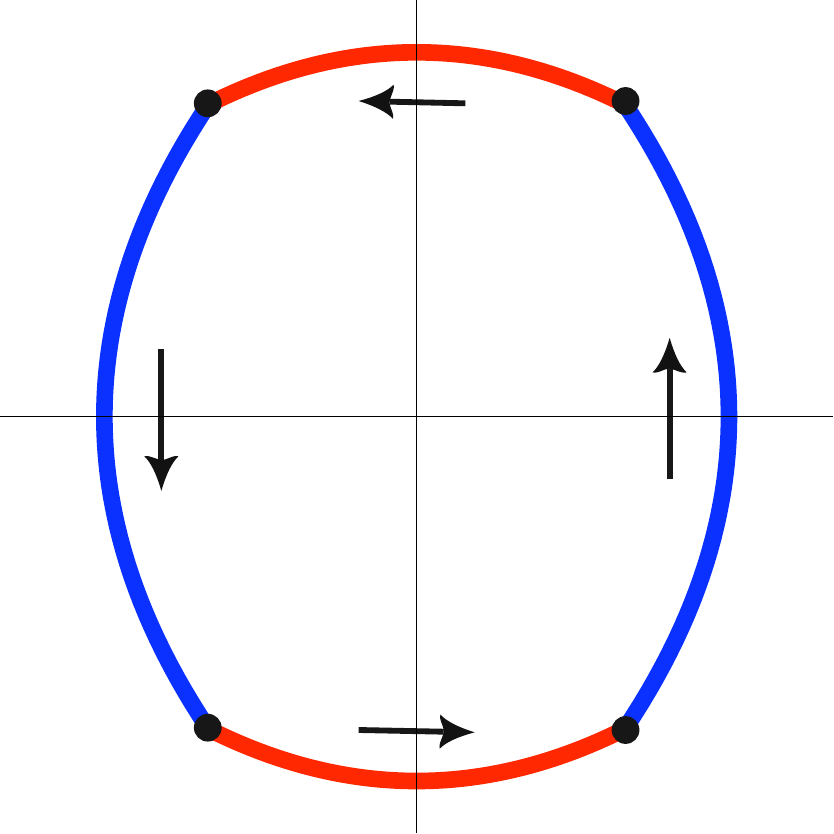}\quad
\includegraphics[scale=0.5]{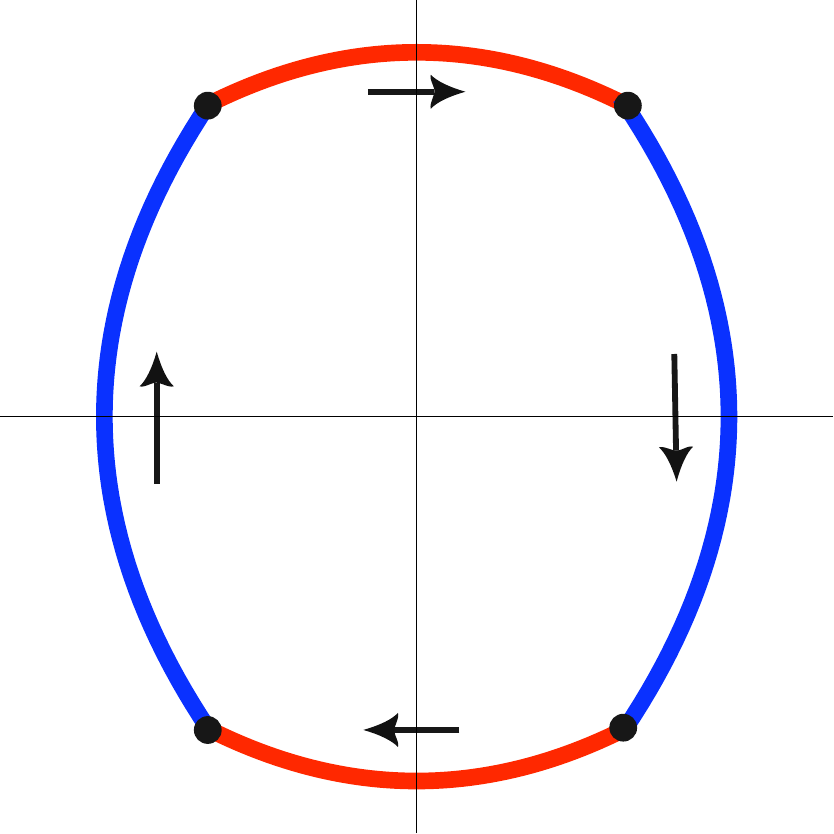}
\caption{Sketch of the four different complex instantons for the field in (\ref{2pairgauge}), which has four complex turning points in the complex $x^0$ plane, shown as solid circles. In the first two figures, the instanton goes from a turning point to its complex conjugate and back again, while in the last two figures the closed trajectory is composed of both instanton and interference trajectories. The difference between the last two is the sense of propagation around the loop.
}
\label{fig4}
\end{figure}

The total contribution to the imaginary part of the effective action is a sum of various types of closed trajectories, composed of both instanton and interference segments. It is best illustrated by the example in Figure \ref{fig4}, again for the field $A_3(x^0)=E/\omega/(1+ (\omega x^0)^2)$, which has two pairs of complex conjugate turning points. We can form closed instanton trajectories by going from one turning point to its complex conjugate and then back again. This is the usual textbook \cite{landau} instanton contribution $e^{- |W_{\rm cl}|}$ to ${\rm Im}\,\Gamma$, corresponding to the first two plots in Figure\ref{fig4}. There are two such contributions because there are two such instantons, one for each complex conjugate pair. [It is an (intentional) accident of this example that the two instantons give exactly the same contribution -- for other fields with two separate instantons, one may have a dominant $e^{- |W_{\rm cl}^{(j)}|}$ factor.]. But there are other closed trajectories composed of {\it both} instanton and interference terms, forming a closed loop through all four turning points, as shown in the last two plots in Figure \ref{fig4} . The sum over closed trajectories in (\ref{approx3}, \ref{approx4}) corresponds in this case to the sum over the four distinct closed trajectories shown in Figure \ref{fig4}. The generalization to fields with more pairs of turning points is clear, and follows the same pattern as the WKB approach in \cite{dd}. Note that closed trajectories consisting of an {\it interference} trajectory and its reverse do not contribute to the imaginary part of the effective action.

For spinor QED, the only change is that we must include the determinant factors appearing in (\ref{approx4}). For this class of fields, these determinant factors reduce to factors of $\pm 1$. This is because
we can write the spin factor as \cite{wli}
\begin{equation}
{\rm tr}\left[e^{i\int  \frac{1}{2}\sigma_{\mu\nu}F^{\mu\nu}}\right]=4\cos\left(\frac{i}{2}\int_0^{T_c}\partial_{0}A_3(x_0)d u\right)
\end{equation}
The integral around the closed loop can be separated into instanton and interference segments, and one finds that
\begin{eqnarray}
\frac{i}{2}\int_{\rm segment} \partial_{0}A_3(x_0)du =\frac{\pi}{2}
\label{segment}
\end{eqnarray}
For an instanton segment this follows from a substitution $y=i\left(p_3+\frac{e}{c}A_3(x^0)\right)$, while for an interference segment we substitute $y=\left(p_3+\frac{e}{c}A_3(x^0)\right)$. This gives a net result of $\cos(m\,\pi/2)$ for a closed trajectory with $m$ segments. Thus, for example in the first two trajectories of Figure \ref{fig4} there are two segments, so there is a spin factor minus sign, which combines with an overall factor of $(-1)$ to give a positive imaginary part of the spinor effective action. On the other hand, for the latter two trajectories shown in Figure \ref{fig4} there are four segments, so there is a positive spin factor, and so these trajectories contribute with the opposite sign. For closed trajectories containing $2m$  segments we get a net spin factor of $\cos(m\pi)$.  This, along with the global sign difference between scalar and spinor QED is  a reflection of the role of quantum statistics in determining the sign of quantum interference terms, and the pattern of signs agrees with the pattern found in the WKB treatment of quantum interference in vacuum pair production \cite{dd}.

\section{Quantitative results}

\subsection{One dominant  pair of turning points}

\begin{figure}[htb]
\begin{center}$
\begin{array}{cc}
\includegraphics[scale=0.6]{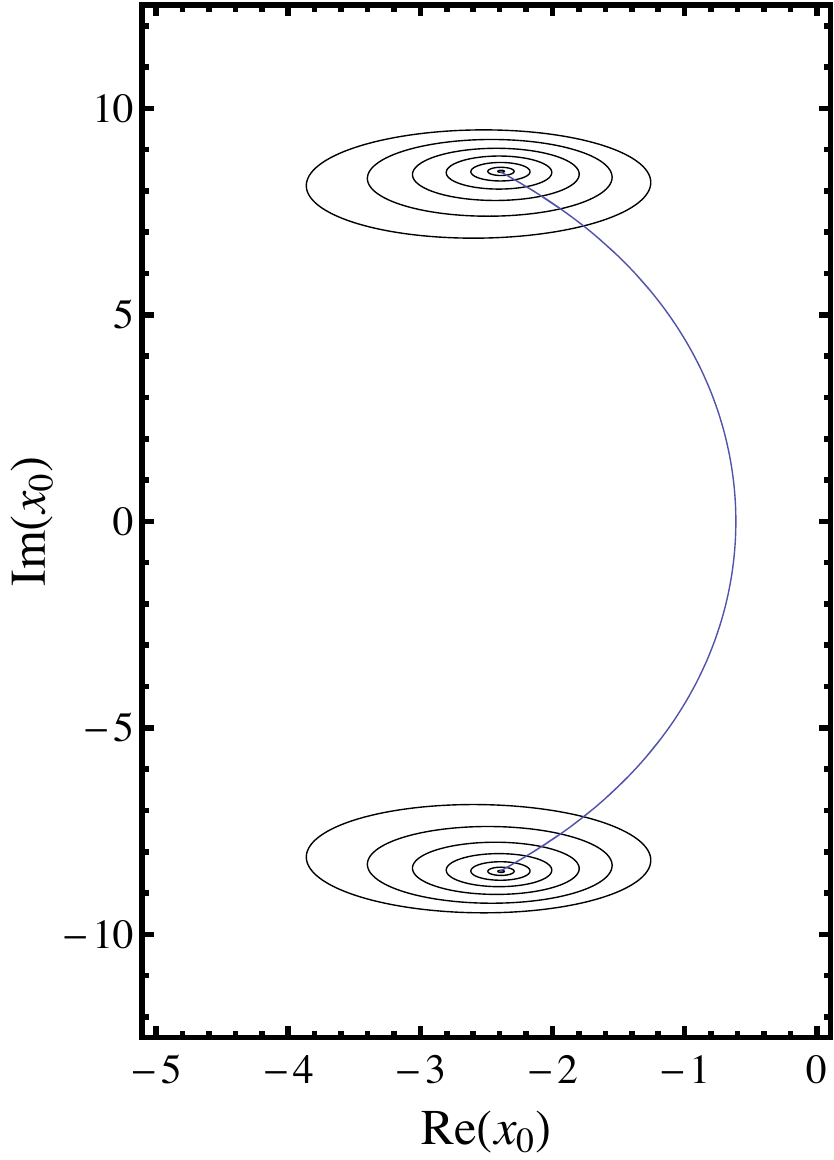}\hskip 3cm
\includegraphics[scale=0.7]{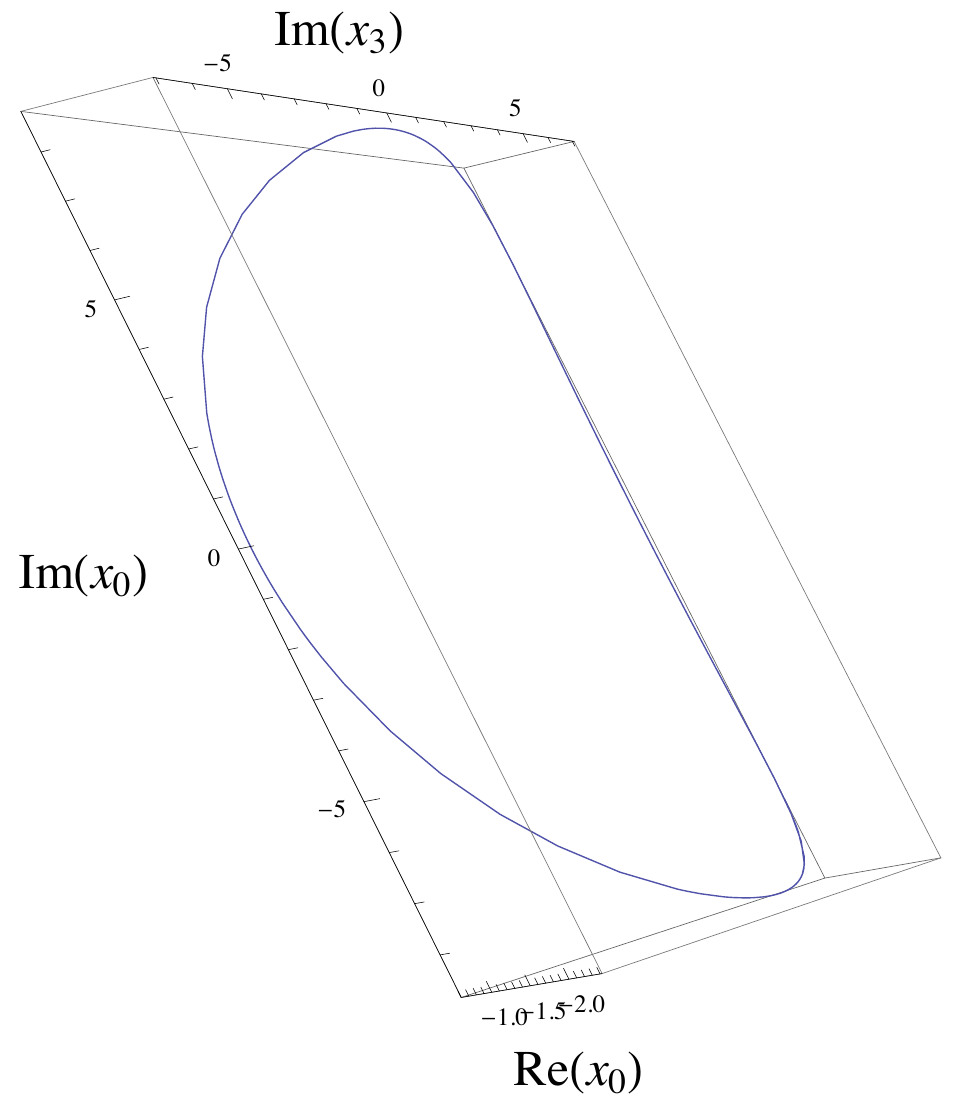}
\end{array}$
\end{center}
\caption{The first plot shows the complex instanton trajectory in the complex $x^0$ plane for the single-pulse electric field in (\ref{1pairgauge}). The trajectory goes from one turning point to its complex conjugate and then back again. The second plot
shows a 3D representation of the same trajectory. The field parameters are: $E_0=0.1$, $\omega=0.1$, and $p_3=0.5$, all in units set by the electron mass scale $m$.}
\label{fig5}
\end{figure}

\begin{figure}[htb]
\begin{center}
\includegraphics[scale=0.8]{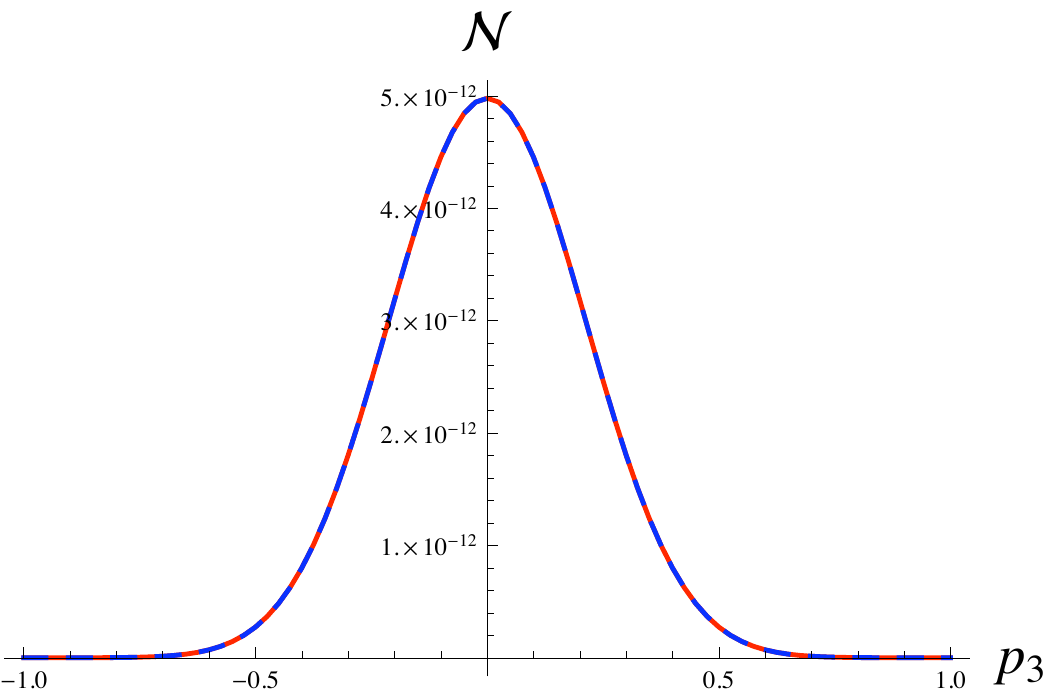}
\end{center}
\caption{The expected particle number, as a  function of longitudinal momentum $p_3$, for both scalar and spinor QED. The
dashed (red) line shows  result (\ref{np1}) evaluated on the complex worldline instanton trajectory, while and
 thick (blue) line shows the WKB result from \cite{dd}. The field parameters are: $E_0=0.1$, and $\omega=0.1$, in units set by the electron mass scale $m$.}
\label{fig6}
\end{figure}

In this section we compute the momentum spectrum of pairs created by a potential with just one dominant pair of turning points. We choose the common example of an exactly soluble gauge field:
\begin{equation}
A(t)=-\frac{E_{0}}{\omega}\,\tanh\left(\omega t\right)\qquad,\qquad E(t)=E_{0}\,{\rm sech}^2\left(\omega t\right)
\label{1pairgauge}
\end{equation}
While this gauge field yields an infinite number of turning points in the complex plane,  the semiclassical amplitude is dominated by one single pair of turning points lying closest to the real $x^0$ axis \cite{dd}.
There is therefore no quantum interference, and
neglecting prefactors, the expectation value of the number of pairs produced with longitudinal momentum $p_3$ follows from (\ref{approx3}, \ref{approx4})
as
\begin{eqnarray}
{\mathcal N}^{\rm scalar}(p_3)\approx {\mathcal N}^{\rm spinor}(p_3)\approx e^{- i\, W_{\rm instanton}(p_3)}
\label{np1}
\end{eqnarray}
where $W_{\rm instanton}$ is evaluated using (\ref{reduced}), and $iW_{\rm instanton}(p_3)$ is real and positive.
When $p_3=0$ there are simple explicit formulae for the the closed trajectories \cite{wli}, but when $p_3\neq 0$ the complex classical closed trajectories are given in terms of inverse elliptic functions, which are cumbersome. Instead, we use a direct numerical integration, for each $p_3$, of the classical equations of motion (\ref{x0-instanton-a}, \ref{x0-instanton-b}, \ref{x0-instanton-c}). A typical trajectory is shown in Figure \ref{fig5}. We then compute the classical action $W_{\rm instanton}$ in (\ref{reduced}), evaluated  on this classical solution. The resulting approximate expression (\ref{np1}) for the particle number is plotted in Figure \ref{fig6} as a function of the longitudinal momentum $p_3$, showing excellent agreement with other methods
%the result obtained from the quantum kinetic equation or the Riccati approach 
(as discussed in \cite{dd}).

\subsection{Two pairs of turning points}

\begin{figure}[htb]
\begin{center}$
\begin{array}{cc}
\includegraphics[scale=0.65]{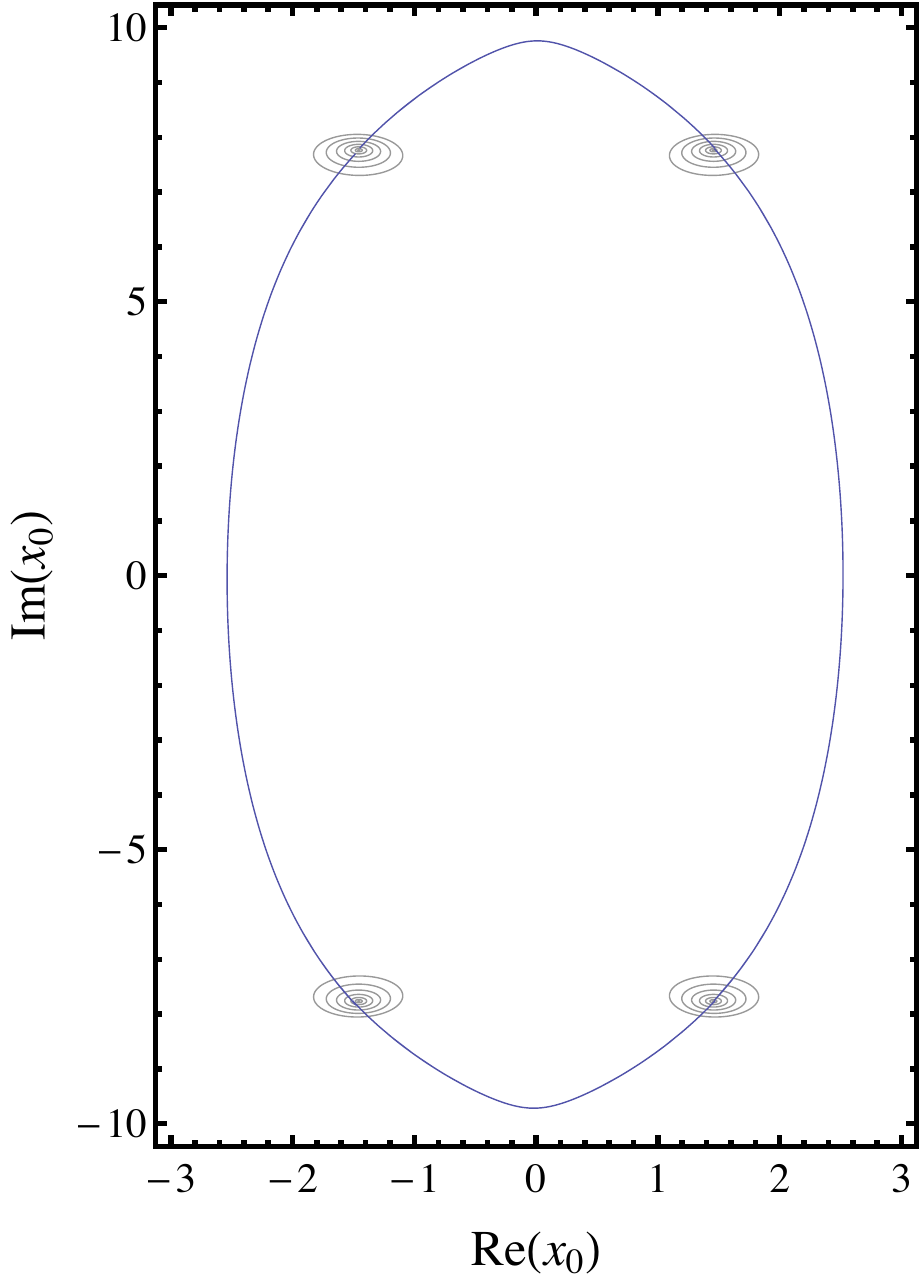}\hskip 2cm
\includegraphics[scale=1]{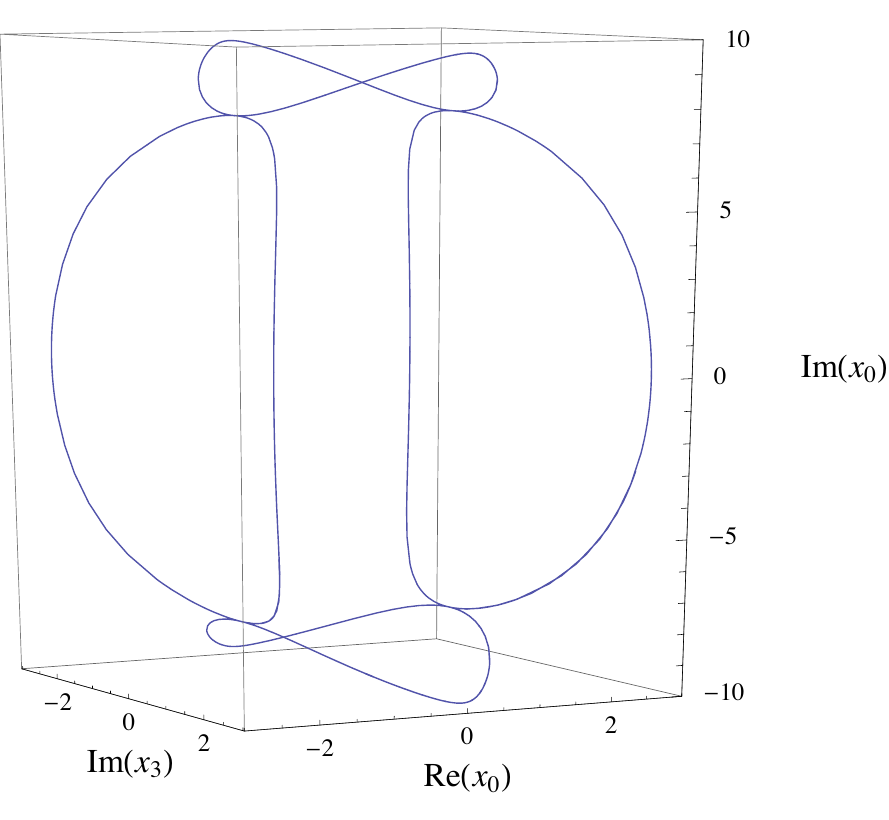}
\end{array}$
\end{center}
\caption{The first plot shows a closed trajectory, in the complex $x^0$ plane, with both instanton and interference segments. The second plot
shows a 3D representation of the same trajectory. The field parameters are: $E_0=0.1$, $\omega=0.1$, and $p_3=1.85$, all in units set by the electron mass scale $m$. }
\label{fig7}
\end{figure}

A more interesting example is provided by a field that induces quantum interference amongst the produced particles. As in \cite{dd}, we consider the field
\begin{equation}
A(t)=\frac{E_{0}}{\omega}\frac{1}{1+\omega^2t^2}\qquad , \qquad E(t)=\frac{2 E_{0}\,\omega\, t}{\left(1+\omega^2t^2\right)^2}
\label{2pairgauge}
\end{equation}
which has precisely two pairs of (complex conjugate) turning points in the complex $x^0$ plane. For each value of longitudinal momentum $p_3$, we integrate the classical equations of motion with their initial conditions, as given in both (\ref{x0-instanton-a}, \ref{x0-instanton-b}, \ref{x0-instanton-c}) and in (\ref{x0-interference-a}, \ref{x0-interference-b}, \ref{x0-interference-c}). This procedure produces four different types of closed trajectories, of the form sketched in Figure \ref{fig4}. A numerical example of an interference trajectory is shown in Figure \ref{fig7}. For each such trajectory, we then compute the classical action $W_{\rm instanton}$ in (\ref{reduced}), evaluated  on this classical solution. As $p_3$ varies, the turning points move, so each trajectory also changes, as does the classical action.
For the instanton-type trajectories, $iW_{\rm instanton}$ is real and positive, while for the interference trajectories $iW_{\rm instanton}$ has both real and imaginary parts, coming from the instanton and interference segments respectively. From (\ref{approx3}) and (\ref{approx4}) we obtain the the semiclassical approximations:
\begin{eqnarray}
{\mathcal N}^{\rm scalar}(p_3)\approx e^{- i\, W_{\rm instanton}^{(1)}(p_3)}+e^{- i\, W_{\rm instanton}^{(2)}(p_3)}
+e^{- i\, W_{\rm instanton}^{(3)}(p_3)}+e^{- i\, W_{\rm instanton}^{(4)}(p_3)}
\label{scalar2}\\
{\mathcal N}^{\rm spinor}(p_3)\approx e^{- i\, W_{\rm instanton}^{(1)}(p_3)}+e^{- i\, W_{\rm instanton}^{(2)}(p_3)}
-e^{- i\, W_{\rm instanton}^{(3)}(p_3)}-e^{- i\, W_{\rm instanton}^{(4)}(p_3)}
\label{spinor2}
\end{eqnarray}
where the superscripts label the trajectory type as shown in Figure \ref{fig4}. Note that for the spinor case the determinant factors in (\ref{approx4}) are +1 for the purely instanton closed trajectories, but are equal to -1 for the interference trajectories, as explained in the previous section and also confirmed numerically. This encodes the difference between spinor and scalar QED in the semiclassical worldline instanton approximation.

\begin{figure}[htb]
\begin{center}
\includegraphics[scale=0.7]{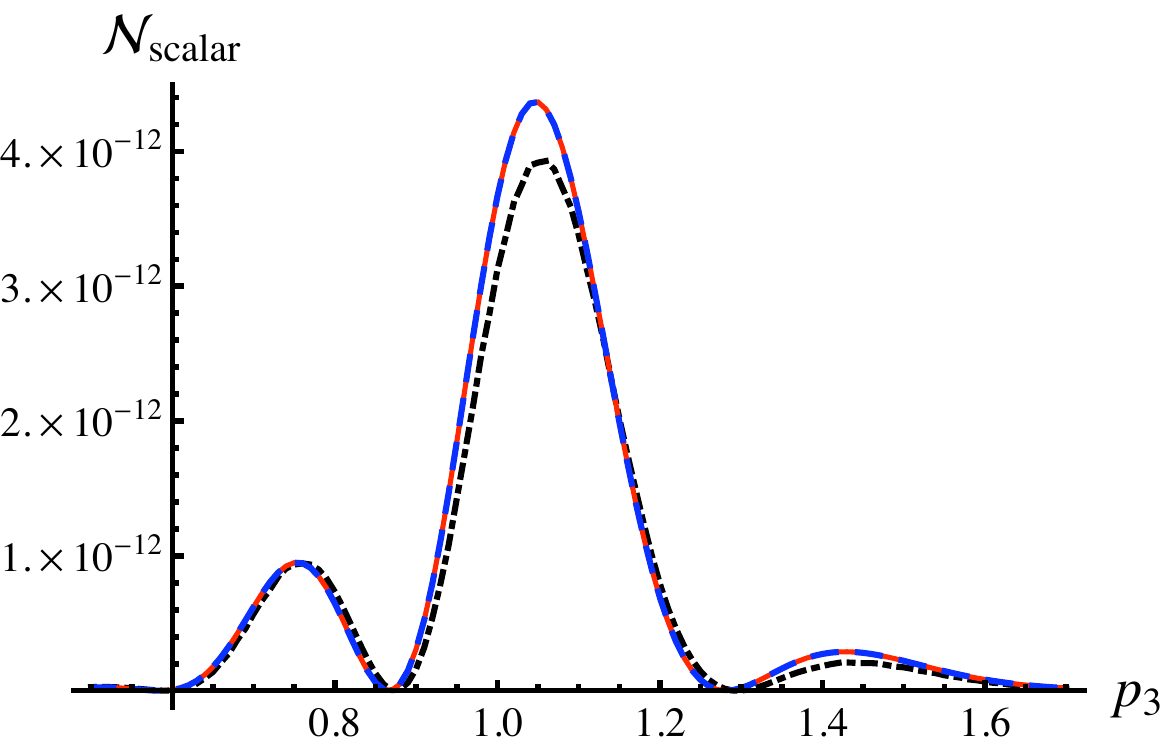}\hskip 1cm
\includegraphics[scale=0.7]{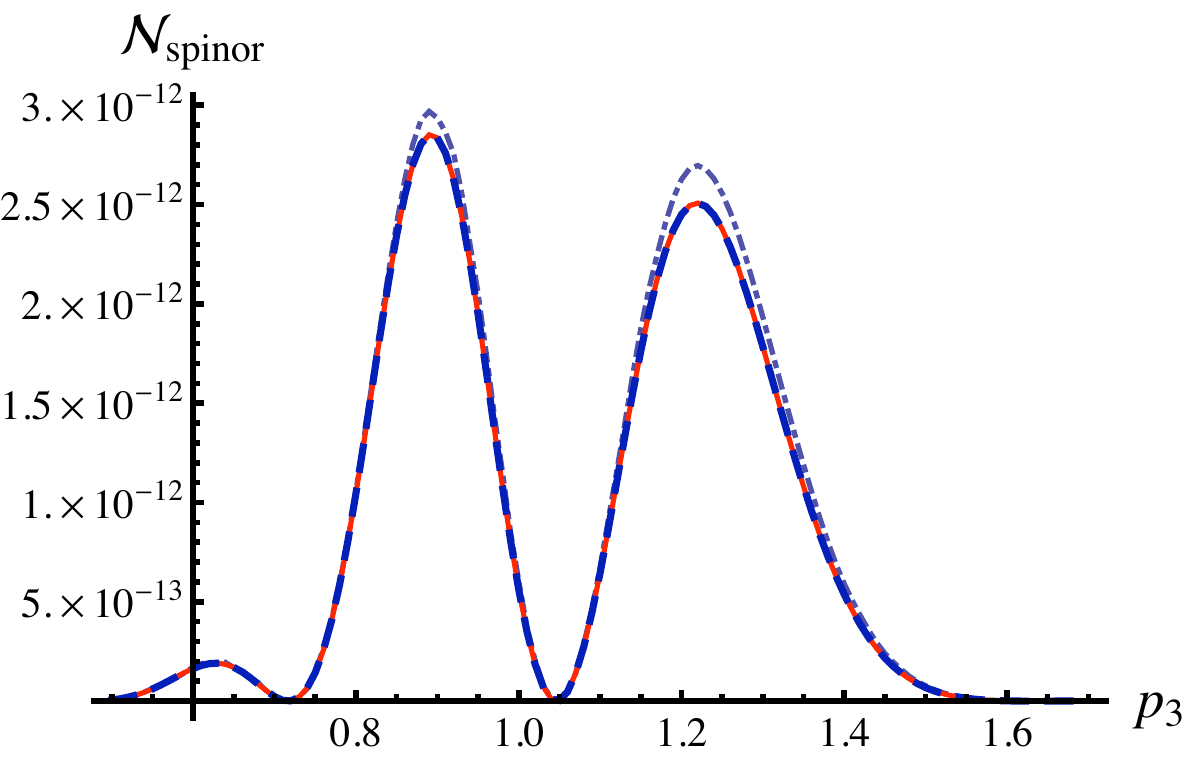}
\end{center}
\caption{
The expected particle number, as a  function of longitudinal momentum $p_3$, for both scalar and spinor QED. The dot-dashed (black) line shows the result of an exact numerical computation. The solid (red) line shows the result (\ref{np1}) evaluated on the complex worldline instanton trajectory, while and
dashed (blue) line shows the WKB result from \cite{dd}. The field parameters are: $E_0=0.1$, and $\omega=0.1$, in units set by the electron mass scale $m$. The first plot is for spinor QED, and the second for scalar QED.}
\label{fig8}
\end{figure}

In fact, for this particular example, the different instanton segments are symmetrical, so that $iW^{(1)}=iW^{(2)}={\rm Re}(i W^{(3)})={\rm Re}(i W^{(4)})$, and ${\rm Im}(iW^{(3)})=-{\rm Im}(i W^{(4)})$. Therefore, in this case we can write
\begin{eqnarray}
{\mathcal N}^{\rm scalar}(p_3)\approx 4 \,e^{- i\, W_{\rm instanton}^{(1)}(p_3)}\,\cos^2\left({\rm Im}\left( \frac{i}{2}\, W_{\rm instanton}^{(3)}(p_3)\right)\right)
\label{scalar2b}\\
{\mathcal N}^{\rm spinor}(p_3)\approx 4\, e^{- i\, W_{\rm instanton}^{(1)}(p_3)}\,\sin^2\left({\rm Im}\left( \frac{i}{2}\, W_{\rm instanton}^{(3)}(p_3)\right)\right)
\label{spinor2b}
\end{eqnarray}
These expressions are plotted in Figure \ref{fig8}, as functions of the longitudinal momentum $p_3$, showing excellent agreement with the exact numerical result and with the WKB results from \cite{dd}.

In the course of this analysis we found an interesting numerical instability that arises for certain values of $p_3$, when the classical trajectories approach poles in the complex plane [which typically arise for localized fields]. In the appendix we present  a simple modification to the numerical procedure that avoids this instability, taking advantage of the einbein formulation of the worldline effective action, due to reparameterization invariance of the associated path integral.

\section{Conclusions}

To conclude, we have shown that it is necessary to consider complex classical trajectories in the semiclassical approximation of the worldline path integral expression for the QED effective action, in order to obtain the pair production probability and the associated momentum spectrum of the produced pairs. This is somewhat surprising, because the original path integral is of course a sum over real closed trajectories in Minkowski spacetime. To extract the momentum dependence and to incorporate the effects of quantum interference it is not sufficient to consider classical solutions in Euclidean spacetime (where time is pure imaginary); one must consider genuinely complex classical solutions.
The semiclassically relevant closed classical trajectories, the "worldline instantons", are composed of segments that we have called "instanton segments", for which the proper-time parameter has constant real part, and "interference segments", for which the proper-time parameter has constant imaginary part. Generically, all these closed trajectories are complex, and so do not arise as solutions to the Euclidean classical equations of motion. With just a single instanton segment this is precisely the instanton prescription of Rubakov et al \cite{rubakov}. If there is more than one instanton solution, then in addition we need to include the interference segments that tie the instanton segments together. These interference trajectories, as shown in the last two plots in Figure \ref{fig4}, produce classical actions $iW$ that have both real and imaginary parts, with the imaginary parts characterizing the quantum interference effects. For spinor QED there are additional determinant factors that give additional minus signs for interference terms. This is how the effect of quantum statistics enters the semiclassical worldline instanton approximation. The agreement with the WKB analysis of \cite{dd}, which in turn agrees very well with exact results, is extremely good.

We end with some brief comments about lessons from this work for the more general case, when the gauge field background represents a more complicated laser configuration, for example colliding short-pulse beams with spatial focussing. In this case, the old-style WKB approaches are not directly useful, but a possible semiclassical approach is provided by the worldline formalism. As emphasized already by Fock, Nambu, Feynman and Schwinger \cite{fock,nambu,feynman,schwinger}, the problem reduces to a problem of tunneling in four-dimensional Minkowski spacetime. As this work has shown, we need to consider complex classical paths, and furthermore the trajectories contain both instanton segments and interference segments. It would be interesting to investigate the applicability of various numerical and analytical methods that have been developed for multi-dimensional tunneling, in the context of non-relativistic two- and three-dimensional problems in chemical and molecular physics \cite{miller,nakamura,ayers}. These methods also typically involve complex trajectories, in the spatial coordinates, and analytic continuation of the ``time'' coordinate to imaginary values. Such an extension would have to take into account the derivative coupling inherent to a gauge theory, whereas most previous results are for Hamiltonians of the form $H=\frac{\vec{p}^2}{2m}+V(\vec{x})$. (In the non-relativistic quantum mechanical context the analogue would correspond to tunneling in the presence of a magnetic field, as has been studied by Dykman \cite{dykman}.) The QED extension would also have to take into account the relativistic causality features of Minkowski space, which result in the very different physical role played by electric and magnetic components of the background field. An interesting class of fields to investigate would be  finite-plane-wave fields \cite{Heinzl:2010vg}, which involve both space and time, and which exhibit analogous quantum interference effects.

\bigskip

\bigskip

We  acknowledge support from the US DOE  grant DE-FG02-92ER40716.

\section{Appendix: an important numerical technicality}

For fields with localized structures, there arises an interesting numerical technicality due to the fact that in addition to turning points in the complex $x^0$ plane, there can also be poles of the function $A_3(x^0)$ in the complex $x^0$ plane. For example, for the field $A_3(x^0)=E/\omega \tanh(\omega x^0)$ these poles lie along the imaginary axis, at $x^0=(n+1/2) \pi\, i/\omega$, while for the field $A_3(x^0)=E/\omega/(1+ (\omega x^0)^2)$ there are precisely two poles, at $x^0=\pm i/\omega$. The locations of these poles do not depend on $p_3$. However, as $p_3$ is varied, we have observed that beyond certain values of $p_3$ the turning points can move into positions where the classical trajectories, obtained by the numerical method outlined in Section IIIC,  come close to the poles. In these cases, when the trajectory passes very close to the pole, the numerical integration of the equation of motion can lead to the trajectory "jumping" to another branch, so that it does not in fact continue to the expected final turning point. This applies to both the instanton and interference trajectories, depending on the location of the turning points relative to the poles.

Similar numerical instabilities [although for slightly different reasons] have been observed in semiclassical  studies of multi-dimensional quantum mechanical problems, where complex instantons are also required \cite{bezrukov}. One possible solution is the prescription of Rubakov and collaborators:  in fact, one only needs to minimize the {\it imaginary part} of the action in the path integral, which in turn means that the appropriate initial condition for instanton segments is to take just the imaginary part of the initial velocity to vanish. This amounts to a stationary phase approximation. Then the real part of the initial velocity is a free parameter, and one should vary with respect to this parameter in order to find those trajectories that give a minimum imaginary part of the classical action. Implementing this prescription in our gauge theory case we find that in the situation where the classical trajectories do not approach the poles, our prescription produces completely equivalent results for the classical action on the trajectory segments. When the longitudinal momentum reaches a threshold value beyond which the trajectories "jump" to another branch, we can cure the situation by only fixing the imaginary part of the initial velocity to vanish. By adjusting the real part of the initial velocity, we can force the trajectory to go to the expected final turning point, and the correct {\it minimum} imaginary action is obtained by tuning the real part of the initial velocity to the threshold value where the jumping is first avoided.
While this procedure can indeed be implemented, in these QED problems we found it also to be extremely delicate numerically. This is because one needs to compute the proper-time interval $T$ of the trajectory segment very precisely in order to evaluate the classical action accurately. A small error in the period $T$ can produce a significant error in the classical action, $iW_{\rm cl}$, because the classical Lagrangian changes sign rapidly in the neighborhood of $T$.

We have found another, more numerically stable, way to overcome the problem of classical trajectories approaching poles. The key observation is that we can relax the proper-time normalization condition (\ref{proper}), by taking advantage of the well-known reparametrization invariance of the path integral \cite{polyakov,kleinert}. Thus, for scalar QED, instead of (\ref{scalar}), we have the more general expression:
\begin{equation}
\Gamma_{\rm eff}^{\rm scalar}[A]=-i\int\mathcal{D}n\,\Phi(n)\int_0^{\infty}\frac{d T}{T}\,e^{-i\frac{m^2 c^2}{2}\int_0^T n(u)du}
\int_{x(0)=x(T)}d^4x\int\mathcal{D}^4x\,e^{-iS}
\label{effaction}
\end{equation}
where $n$ represents the auxiliary einbein field, which is to be fixed with the aid of the $\Phi(n)$ "gauge-fixing"  functional \cite{polyakov,kleinert}.  The action including the einbein field is:
\begin{equation}
S[x^\mu(\tau)]=\int_0^T \left(\frac{1}{2 n}\frac{dx_\mu}{du}\frac{dx_\mu}{du} -\frac{e}{c} \dot{x}_{\mu}A^{\mu}(x)\right) du
\label{action-einbein}
\end{equation}
Previously we chose the "gauge-fixing" condition $n=1$,  but we are free to rescale proper-time by any factor.
 In particular, note that our choice of imaginary $u$ for the instanton segments can be thought of as taking $n=i$.  
 %Thus, this einbein freedom is related to deformations of the $T$ contour.
 \begin{figure}[htb]
\begin{center}$
\begin{array}{cc}
\includegraphics[scale=0.7]{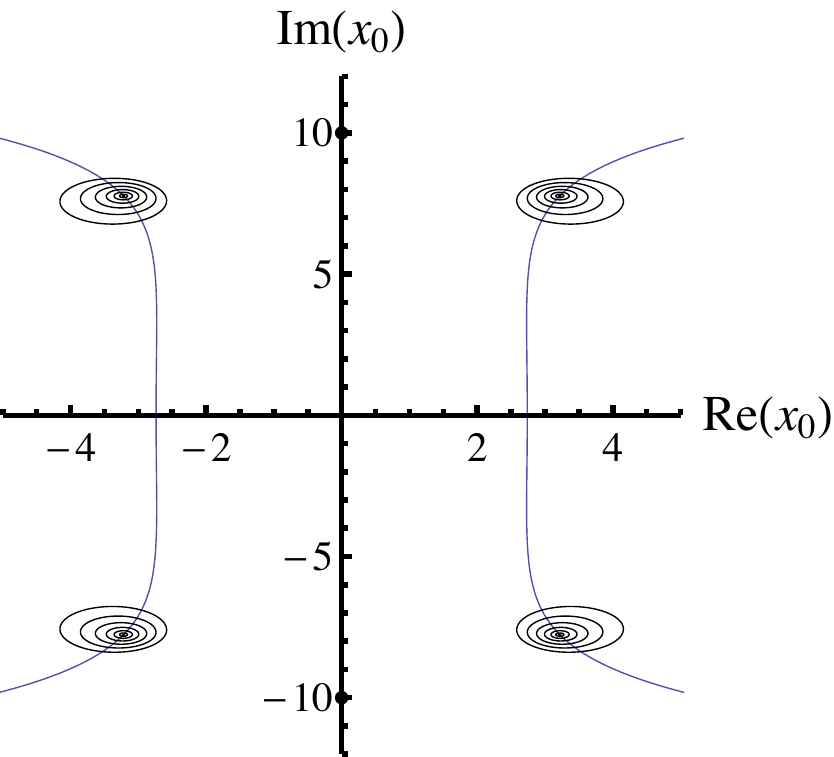}\hskip 3cm
\includegraphics[scale=0.7]{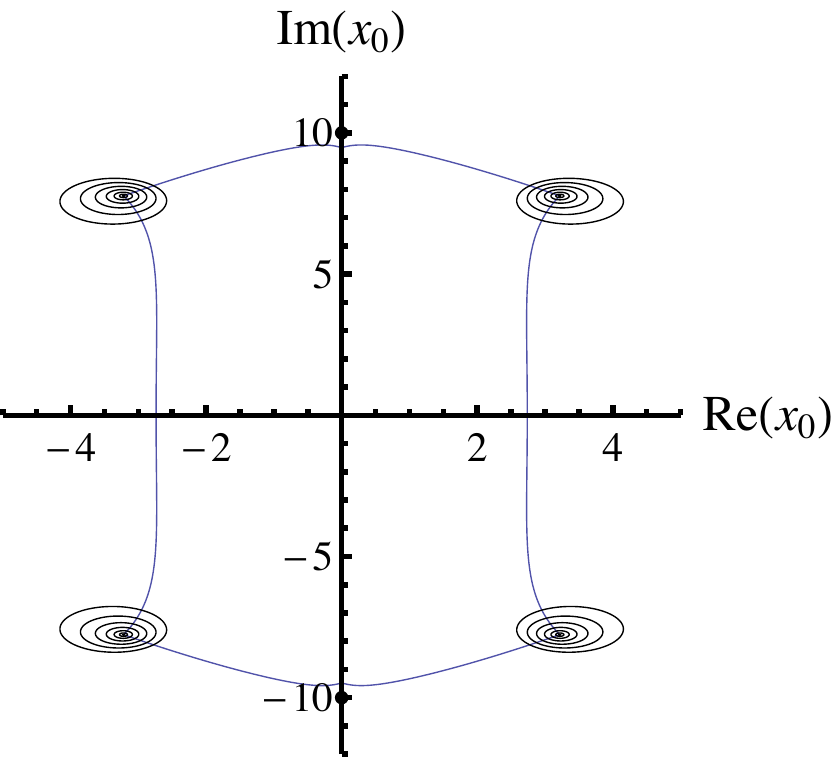}
\end{array}$
\end{center}
\caption{Producing closed trajectories by tuning the einbein $n$, as explained in the text. These plots are for the two-turning-point-pair gauge field in (\ref{2pairgauge}). The first plot shows the situation for $E_0=0.1$, $\omega=0.1$, and $p_3=1$. The interference trajectories do not connect the different turning point pairs, due to the appearance of poles at $x^0=\pm i/\omega=\pm 10i$. The second plot shows the result of choosing  $n=1\pm 0.74i$ and integrating the modified equations (\ref{einbein-eqs}). Now the poles are avoided and a closed classical trajectory results. The action (\ref{reduced-n}) evaluated on this trajectory gives the correct semiclassical approximation to the particle spectrum, as plotted in Figure \ref{fig8}.}
\label{fig9}
\end{figure}

 Including this einbein factor, the classical equations of motion, together with the constant of integration, read:
 \begin{eqnarray}
 \frac{d^2x^\mu}{du^2}=n\, \frac{e}{c}F^{\mu\nu}(x)\,\frac{dx_\nu}{du}\qquad, \qquad
\frac{dx^\mu}{du}\frac{dx_\mu}{du}=n^2 m^2 c^2
\label{eom-n}
\end{eqnarray}
Therefore, instead of numerically integrating the equations of motion (\ref{x0-instanton-a}, \ref{x0-instanton-b}, \ref{x0-instanton-c}) and (\ref{x0-interference-a}, \ref{x0-interference-b}, \ref{x0-interference-c}), we can integrate the following coupled equations, with associated initial conditions:
\begin{eqnarray}
\frac{d^2x^0}{du^2}&=&-n\,\frac{e}{c} \frac{dx_3}{du}\,\partial_0 A_3(x^0)
\nonumber\\
\frac{d^2x_3}{du^2}&=&-n\,\frac{e}{c} \frac{dx^0}{du}\, \partial_0 A_3(x^0)
\nonumber\\
x^0(u_{\rm initial})&=&x^0_{(j)} \qquad , \quad \text{a turning point solution of (\ref{tps})}
\nonumber\\
x_3(u_{\rm initial})&=&0
\nonumber\\
\left. \frac{dx^0}{du}\right |_{u=u_{\rm initial}}&=&0
\nonumber\\
\left. \frac{dx_3}{du}\right |_{u=u_{\rm initial}}&=&-i \,n\,\,m\,c
\label{einbein-eqs}
\end{eqnarray}
The choice $n=1$ gives the interference segments from before, while the choice $n=i$ gives the instanton segments as before. Now, to find the instanton segments we choose $n=i(1+i b)$, and for the interference segments we choose $n=(1+i b)$, for some real $b$. We can tune $b$ so that the trajectories avoid the poles. Finally, given such a trajectory, we compute the associated classical action as
\begin{eqnarray}
W[x^0(u); \frac{1}{2}m^2c^2]= \int_0^T \frac{1}{n}\,\left(\frac{dx^0}{du}\right)^2 du
%-\frac{1}{2} mc^2\right)
\label{reduced-n}
\end{eqnarray}
As an illustration of this procedure, consider the field (\ref{2pairgauge}) with two sets of complex conjugate turning points. In the first plot in Figure \ref{fig9} we see that for $p_3=3$ the interference trajectories do not connect the different pairs of turning points, as expected, but go off to infinity.  This is with the choice $n=1$. We can cure this by tuning $n$ to take the value $n=1\pm 0.74 i$, with the resulting trajectories shown in the second plot of Figure \ref{fig9}. This choice of $n$ is chosen so that the classical trajectories pass safely by the poles that occur at $x^0=\pm i/\omega$ [in the plot, $1/\omega=10$]. They are patched together smoothly on the imaginary $x^0$ axis. 
Now we have the desired instanton segments, and after evaluating the classical action (\ref{reduced-n}), we obtain the correct semiclassical contribution to the particle number momentum spectrum, as shown in Figure \ref{fig8}. Note that for this $p_3=3$, and $n=i$, the instanton segments do indeed connect the complex conjugate turning points.
We have found that by suitable tuning of the einbein factor $n$ we are always able to avoid the poles and produce closed classical trajectories that avoid the poles.

\end{document}